\documentclass{article}

\usepackage{amsfonts}

\def\mb{\mathbb}

\def\bea{\begin{eqnarray}}
\def\eea{\end{eqnarray}}
\def\({\left(}
\def\){\right)}
\def\<{\left<}
\def\>{\right>}

\def\tr{{\mbox{tr}}}
\def\bea{\begin{eqnarray*}}
\def\eea{\end{eqnarray*}}
\def\ben{\begin{eqnarray}}
\def\een{\end{eqnarray}}
\def\({\left(}
\def\){\right)}
\def\<{\left<}
\def\>{\right>}

\def\[{\left[}
\def\]{\right]}

\def\+{\bar}
\def\mb{\mathbb}
\def\tr{{\mbox{tr}}}

\def\L{{\cal{L}}}
\def\t{\widetilde}

\def\l{\ell}

\def\A{{\cal{A}}}
\def\B{{\cal{B}}}
\def\M{{\cal{M}}}
\def\N{{\cal{N}}}
\def\O{{\cal{O}}}
\def\U{{\cal{U}}}
\def\V{{\cal{V}}}
\def\W{{\cal{W}}}

\def\X{{\cal{X}}}
\def\Y{{\cal{Y}}}

\def\T{{\cal{T}}}
\def\G{{\cal{G}}}

\begin{document}
\setlength{\unitlength}{1mm}

\pagestyle{empty}
\vskip-10pt
\vskip-10pt
\hfill 
\begin{center}
\vskip 3truecm
{\Large \bf
An associative star-three-product\\
and \\
applications to M two/M five-brane theory}\\ 
\vskip 2truecm
{\large \bf
Andreas Gustavsson\footnote{a.r.gustavsson@swipnet.se}}\\
\vskip 1truecm
{\it  Center for quantum spacetime (CQUeST), Sogang University, Seoul 121-742, Korea}\\
and\\
{\it School of Physics \& Astronomy, Seoul National University, Seoul 151-747 Korea}
\vskip 2truecm
\end{center}
{\abstract{The star-product between functions enable us to take the large $N$ limit in a controlled way. At finite $N$ it serves as a substitute for matrix multiplications. Non-abelian gauge theory can be deconstructed from lower dimensional gauge theories using star-products. In this paper we extend the star-product to a star-three-product. We then apply the star-three-product to realize hermitian three-algebra of ABJM theory. We define a fuzzy three-torus. We deconstruct Abelian M five-brane in a constant background three-form potential on a fuzzy three-torus. We deconstruct non-Abelian extensions which might be related with multiple M five branes. We also mention the fuzzy three-sphere case.}}

\vfill 
\vskip4pt
\eject
\pagestyle{plain}
\section{Introduction}

A maximally supersymmetric and $SO(8)$ invariant theory in three dimensions is the BLG theory \cite{Bagger:2007jr}, \cite{Gustavsson:2007vu}. This is a candidating theory for M two-branes, but seems to be associated with an orbifolded target space \cite{Lambert:2010ji}. The BLG theory can be formulated using a real three-algebra in which case the matter fields are valued in the three-algebra. It can also be formulated in terms of its associated Lie algebra in which case the matter fields are bifundamental fields. The only finite-dimensional matrix realization of a real three-algebra corresponds to $SO(4)$ gauge group \cite{Schnabl:2008wj}, \cite{Papadopoulos:2008sk}, \cite{Gauntlett:2008uf}. 

One gets more flexibility if one breaks $SO(8)$ to $SU(4)$, where one finds an infinite class of theories with finite-rank gauge groups. These are the ABJM theories \cite{ABJM}. These theories can be formulated in terms of hermitian three-algebra \cite{Bagger:2008se}. They can also be formulated in terms of its associated Lie algebra in which case the matter fields are bifundamental fields. Despite only $SU(4)$ symmetry is manifest in the classical Lagrangian, the $SO(8)$ is a hidden symmetry when the Chern-Simons level takes the values $k=1,2$ and the gauge group is $U(N)\times U(N)$ \cite{Bashkirov:2010kz}, \cite{Gustavsson:2009pm}. 

A few things remain unclear regarding ABJM theory as viable theory of M two-branes though. Quite little is known about the theory for Chern-Simons level $k=1$ where it is expected to describe M two-branes in flat eleven-dimensional target space. It is not clear how ABJM theory for $k=1$ can be reduced to D2 branes by compactifying one transverse direction. It is not clear how to obtain the $M^{3/2}$ scaling of degrees of freedom on $M$ coincident M two-branes \cite{Henningson:1998gx}, \cite{Klebanov:1996un} for $k=1$. In mass deformed ABJM theory we can not get the fuzzy three-sphere vacuum manifold. We can not deconstruct the M five-brane from ABJM theory. The best that has been done in that direction is to deconstruct D4 brane on a fuzzy two-sphere in the large $k$ limit \cite{Nastase:2010uy}.

We can deconstruct non-Abelian $D+2$ dimensional Yang-Mills theory from $D$ dimensional Yang-Mills matter theory \cite{Iso:2001mg}, \cite{Myers:1999ps}. The idea is to expand $D$ dimensional Yang-Mills theory about a fuzzy two-sphere vacuum in small fluctuations. To get Maxwell theory on $M_{D+2}$ (a classical $D$ dimensional manifold $M_D$ times a fuzzy $S^2$) we can realize the algebra by a finite set of spherical harmonics on $S^2$ and use a star-product to define the  Lie bracket of two functions,
\bea
[f,g] &=& f * g - g * f.
\eea
To get a non-Abelian Yang-Mills theory on $M_{D+2}$, we can promote these functions to matrix-valued functions $f M$ where $M$ is a matrix. The Lie bracket defined as
\bea
[f M , gN] &=& f * g MN - g * f NM
\eea
again satisfies the Jacobi identity thanks to associativity of the star-product. The commutator can be expressed as $[f,g]MN+g*f[M,N]$. The term which is on the form $[f,g]MN$ will be present in the Abelian as well as the non-Abelian theory. The commutator term $g*f[M,N]$ will implement the non-Abelian gauge structure in the deconstructed theory. More precisely the matrices $M$ will generate the non-Abelian gauge group of a super Yang-Mills theory on $M_D \times S^2$. We do not have to assume large $N$ limit of matrices in order to map them into functions when using star-products. But it is true that when eventually taking the large $N$ limit it is advantageous to work with star-products instead of matrix multiplications since it is not easy to visualize matrices of infinte size. The star-product enable us to take the large $N$ limit in a controlled way.

To generalize this to M two-branes and deconstruct M five-branes it is natural to seek an associative star-three-product -- a product that multiplies together three functions. No product is defined, needed or wanted which multiplies together only two functions. 

The M five-brane has no Lagrangian description if one insists on keeping $SO(1,5)$ Lorentz invariance manifest. If we break $SO(1,5)$ we can write down a Lagrangian though. If one deconstructs the M five-brane from M two-branes, the natural way of breaking the Lorentz symmetry is to $SO(1,2)\times SO(3)$. This symmetry breaking was studied in \cite{Pasti:2009xc} following \cite{Ho:2008ve}. The full $SO(1,5)$ Lorentz symmetry is not manifest, but it supposed to be hidden in this formulation.

Given a star-three-product we may attempt to deconstruct non-Abelian M five-brane theory. However the star-three-product has other applications as well. It provides a new way of formulating M two-brane theories, in which $SO(8)$ symmetry is kept manifest in the classical Lagrangian just like in the original BLG theory. Thus the star-three-product circumvents the no-go theorem that says that the only finite-dimensional gauge group that admits an $SO(8)$ invariant Lagrangian is $SO(4)$. As we thus have a consistent truncation to finite-dimensional three-algebras we can also provide a more rigorous way of counting the $M^{3/2}$ degree of freedom, following the idea in \cite{Chu:2008qv}. The counting argument in this reference was incomplete just because there a consistent finite-dimensional truncation was missing. 

In section 2 we recall the hermitian three-algebra. In section 3 we define the star-three-product and in the Appendix $B$ we demonstrate associativity. The goal of the many subsections in section 3, which may seem rather technical, is to connect the functions with matrices, essentially by reducing the three-star-product to a star-product. In section 4 we finally come to physics. We rewrite ABJM Lagrangian in a manifestly $SO(8)$ invariant form. In section 5 we deconstruct the Abelian M five-brane Lagrangian by considering three compact scalar fields compactified on a fuzzy three-torus. We find that the deconstructed Lagrangian can be matched with the expected gauge invariant Abelian M five-brane Lagrangian in a constant background $C$-field that couples to the selfdual part of the field strength on the worldvolume of the M five-brane. In section 6 we deconstruct non-Abelian extensions. In section 6.1 we introduce the concept of a Cartan sub-three-algebra and argue for the $M^{3/2}$ scaling. In section 7 we apply the star-three-product to the GRVV equation to show that this can describe a fuzzy three-sphere.

While this work was in its final stage, two papers \cite{Lambert:2010wm}, \cite{Chen:2010br} appeared which might touch upon a bit similar questions as we address in this paper.

\section{Hermitian three-algebra}\label{Hermitian three-algebra}
Before we can motivate the properties of the star-three-product and eventually present an explicit expression for it, we have to first review hermitian three-algebra. The $\N=6$ supersymmetric ABJM theory  can be set up for any so called hermitian three-algebra $\A$ as shown in \cite{Bagger:2008se}. The three-algebra $\A$ is specified by a three-bracket that closes on three elements of $\A$. That is $(T^a,T^b,T^c) \in \A\times \A \times \A$ is mapped into an element $[T^a,T^b;T^c]\in \A$. To each generator $T^a$ we have a conjugate generator that we denote $T_a$. We have a trace form with the properties 
\bea
\<T^a,T^b\> &=& \<T_b,T_a\>,\cr
\<T^a,T^b\>^* &=& \<T^b,T^a\>,
\eea
which is linear in its first entry and complex anti-linear in its second entry, and which is invariant
\bea
\<[T^a,T^c;T^d],T^b\> - \<T^a,[T^b,T^d;T^c]\> &=& 0.
\eea
The three-bracket satisfies the hermitian fundamental identity
\bea
[[T^a,T^b;T^c],T^e;T^f] &=& [[T^a,T^e;T^f],T^b;T^c] + [T^a,[T^b,T^e;T^f];T^c]\cr
&& - [T^a,T^b,[T^c,T^f;T^e]].
\eea
The three-bracket of a hermitian three-algebra is linear in its two first entries and complex anti-linear in its third entry. It is antisymmetric under exchange of the two elements in the first two entries.

We can now derive the property
\ben
\<[T^a,T^b;T^c],T^d\>^* &=& \<[T^d,T^c;T^b],T^a\>\label{relevant}
\een
using trace invariance. In terms of structure constants, defined as $[T^a,T^b;T^c] = f^{ab}{}_{cd} T^d$ the condition reads\footnote{Anti-linearity implies opposite index placings, so for instance in $[T^a,T^b;T^c] = f^{ab}{}_{cd} T^d$ we have the index $c$ up-stairs inside the three-bracket. When we pull this index outside the anti-linear three-bracket, this index comes down-stairs as seen in the right-hand side. We can not define the three-bracket as $[T^a,T^b;T_c]$ as we do not assume that $T_c$ belongs to the same three-algebra as $T^a$. The three-bracket shall map three elements in the three-algebra into a new element in the three-algebra.}
\bea
f^{*ab}{}_{cd} &=& f^{dc}{}_{ba}.
\eea
but this does not tell us how complex conjugation acts on the three-bracket itself. It is clear that complex conjugation must act on the three-bracket like
\bea
[T^a,T^b;T^c]^* &=& \mu [T_a,T_b;T_c]
\eea
for some $\mu$. The index $c$ must be in the third entry because the result must be antisymmetric in $a$ and $b$. This uniquely fixes this form up to a constant $\mu$. Repeated complex conjugation gives
\bea
[T^a,T^b;T^c]^{**} = (\mu[T_a,T_b;T_c])^* = \mu^* \mu [T^a,T^b;T^c]
\eea
and so 
\bea
\mu^* \mu &=& 1.
\eea
The conjugation of the three-bracket itself has no significance. We can not detect the value of $\mu$ in the Lagrangian. The only relevant property is (\ref{relevant}). We note that under $T^a \rightarrow T^a_{new} = e^{i\varphi} T^a$ the three-bracket transforms the same way, $[T^a_{new},T^b_{new};T^c_{new}] = e^{i\varphi} [T^a,T^b;T^c]$. This implies that 
\bea
[T^a_{new},T^b_{new};T^c_{new}]^* &=& e^{-2i\varphi} \mu [T^a_{new},T^b_{new};T^c_{new}].
\eea
This manifests the insignificance of the phase $\mu$. By a rotation of the generators we can bring the phase $\mu$ into any value.

\subsection{Matrices}
All finite-dimensional realizations of hermitian three-algebra realized by matrices have a three-bracket (or three-commutator) that is given by \cite{Cherkis:2008ha}
\ben
[T^a,T^b;T^c] &=& \lambda(T^a T_c T^b - T^b T_c T^a).\label{finite}
\een
This bracket satisfies the hermitian fundamental identity thanks to the associativity of matrix multiplication. Conversely if the multiplication is not associative, this bracket does not satisfy the fundamental identity. 

The inner product is
\bea
\<T^a,T^b\> &\propto & \tr(T^a T_b)
\eea
where $T_a$ denotes the hermitian conjugate matrix of $T^a$. Demanding trace invariance,
\bea
\tr(\lambda(T^a T_f T^e-T^e T_f T^a)T_b) - \lambda^* \tr(T^a (T_f T^e T_b - T_b T^e T_f)) &=& 0
\eea
we see that 
\bea
\lambda &=& \lambda^*.
\eea
In that case we get
\bea
[T^a,T^b;T^c]^* &=& -[T_a,T_b;T_c]\label{conj}
\eea
with a minus sign.

To set a convenient convention once and for all, we will from now on always set
\bea
\lambda &=& 1.
\eea

In this matrix realization we can think of the matrices $T^a$ as bifundamentals of a product gauge group $G_1 \times G_2$. If the gauge group is $U(N)\times U(M)$ they transform in the representation $(N,\bar{M})$. The matrices $T_a$ will then transform in the representation $(M,\bar{N})$. A product like $T^a T_a$ will transform in $N\times \bar{N} = 1 \oplus $adj of the $U(N)$ factor and as a singlet of the $U(M)$ factor.

\subsection{Functions}
In addition to the finite-dimensional realizations, we have the infinite dimensional realizations  realized by the totally antisymmetric Nambu three-bracket,
\ben
[\T^a,\T^b;\T^c] &=& \hbar*(d\T^a \wedge d\T^b \wedge d\T_c)\label{infinite}
\een
Here $\T^a$ are complex-valued functions on a three-manifold $M$ where $\T_a = \T^{*a}$. The Hodge dual is with respect to the metric on $M$. We must have the complex conjugate generator in the third entry to make the three-bracket complex antilinear in its third entry. We have introduced $\hbar$ which we will identify as a non-three-commutativity parameter. The inner product is defined as
\bea
\<\T^a,\T^b\> &\propto & \int \T^a \wedge *\T_b.
\eea
Trace invariance 
\bea
\hbar\int *(d\T^a \wedge d\T^e \wedge \T^f) \wedge *\T_b - \hbar^*\int \T^a \wedge **(d\T_b \wedge dT_f \wedge dT_e)
\eea
implies
\bea
\hbar &=& \hbar^*
\eea
in which case the above quantity becomes a total derivative
\bea
\hbar \int d(\T^a \T_b d\T^e \wedge \T_f) &=& 0.
\eea
We then get
\bea
[\T^a,\T^b;\T^c]^* &=& [\T_a,\T_b;\T_c]
\eea
with a plus sign. Then $i\T^a$ gives the minus sign. Alternatively we define $\T^{*a} = -\T_a$.

We define the Nambu bracket as
\bea
\{\T^a,\T^b,\T_c\} &=& *(d\T^a \wedge d\T^b \wedge d\T_c)
\eea
If we parametrize $M$ by three local coordinates $\sigma^{\alpha}$ and denote the metric tensor as $g_{\alpha\beta}$, the totally antisymmetric tensor as $\epsilon_{\alpha\beta\gamma}= \pm 1$ and rise all indices by the inverse metric, and denote the measure on $M$ as $d^3 \sigma \sqrt{g}$ then 
\bea
\{\T^a,\T^b,\T_c\} &=& \sqrt{g}\epsilon^{\alpha\beta\gamma}\partial_{\alpha}\T^a\partial_{\beta}\T^b\partial_{\gamma}\T_c.
\eea
The hermitian three-algebra is almost trivially induced by the real three-algbra that is obeyed by the Nambu bracket,
\bea
\{\{\T^a,\T^b,\T_c\},\T^e,\T_f\} &=& \{\{\T^a,\T^e,\T_f\},\T^b,\T_c\} + \{\T^a,\{\T^b,\T^e.\T_f\},\T_c\} \cr
&& + \{\T^a,\T^b,\{\T_c,\T^e,\T_f\}\}
\eea
To see this we just need to rewrite the third term in the right hand side as
\bea
\{\T^a,\T^b,\{\T_c,\T^e,\T_f\}\} &=& -\{\T^a,\T^b,\{\T_c,\T_f,\T^e\}\}
\eea
utilizing the total antisymmetry of the Nambu bracket. 

For the matrix realization we have the Leibniz rule for three-products
\bea
[T^a T_c T^b,T^e;T^f] &=& [T^a,T^e;T^f] T_c T^b + T^a [T_c,T_f;T_e] T^b + T^a T_c [T^b,T^e;T^f].\label{Leibniz}
\eea
but for the Nambu bracket we have a different Leibniz rule,
\bea
\{\T^a \T_c \T^b,\T^e,\T_f\} &=& \{\T^a,\T^e,\T_f\} \T_c \T^b - \T^a \{\T_c,\T_f,\T^e\}\T^b + \T^a \T_c \{\T^b,\T^e,\T_f\}.
\eea
This structure does not carry over to the Nambu bracket. In the next section we will see how to remedy this and other problems with the Nambu bracket by finding the appropriate extension of the Nambu bracket.

\section{Associative star-three-product}
Up to now it seems like we have two different realizations of hermitian three-algebra, one in terms of matrices and one realized by a Nambu bracket. At first glance there is no similarity between these two realizations, and this makes it hard to understand how one can smoothly take a limit of large matrices and map the three-bracket of matrices into something like a Nambu bracket in that limit. To this end we would like to find a map from matrices into functions 
\bea
T^a \rightarrow \T^a
\eea
such that matrix three-multiplication is replaced by star-three-multiplication
\bea
T^a T_c T^b \rightarrow \T^a * \T_c * \T^b
\eea
in such a way that the three-algebra is the same after and before the map. Matrix multiplication is associative
\bea
(T^a T_c T^b) T_e T^f = T^a (T_c T^b T_e) T^f = T^a T_c (T^b T_e T^f)
\eea
Associativity is what makes the matrix realization of the three-bracket (\ref{finite}) satisfy the hermitian fundamental identity. The star-three-product must inherit associativity,
\ben
(\T^a * \T_c * \T^b) * \T_e * \T^f = \T^a * (\T_c * \T^b * \T_e) * \T^f = \T^a * \T_c * (\T^b * \T_e * \T^f).\label{assoc}
\een
The Nambu bracket now finds its natural place as the first order term in the three-bracket that we shall now define as
\ben
[\T^a,\T^b;\T^c] \equiv \T^a * \T_c * \T^b - \T^b * \T_c * \T^a = \hbar \{\T^a,\T^b,\T_c\} + \O(\hbar^2).\label{brackets}
\een
The naive attempt to define the star-three-product as
\bea
(\T^a * \T_c * \T^b)(\sigma) &=& \lim_{\sigma'\rightarrow \sigma} \Big\{ \lim_{\sigma''\rightarrow \sigma'} \exp\(\frac{\hbar}{2}\sqrt{g}\epsilon^{\alpha\beta\gamma}\partial_{\alpha} \partial''_{\beta} \partial'_{\gamma}\)\T^a(\sigma)\T_c(\sigma') \T^b(\sigma'')\Big\}
\eea
which we will also express as
\ben
(\T^a * \T_c * \T^b)(\sigma) &=& \lim_{\sigma'\rightarrow \sigma} \lim_{\sigma''\rightarrow \sigma'} \exp\(\frac{\hbar}{2}\{\bullet,\bullet'',\bullet'\}\)\T^a(\sigma)\T_c(\sigma') \T^b(\sigma'')\label{naive}
\een
is incomplete. If we define the star-product like this, then it will not be associative. To see this let us consider 
\bea
(\T^a * \T_c * \T^b) * \T_e * \T^d 
\eea
At zeroth order in $\hbar$ we have just the usual multiplication of functions which is associative, but at linear order in $\hbar$ we have
\bea
&&\frac{\hbar}{2} \(\{T^a,\T^b,\T_c\}\T^d\T_e + \{\T^a\T_c\T^b,\T^d,\T_e\}\)\cr
&=& \frac{\hbar}{2} \(\{T^a,\T^b,\T_c\}T^d\T_e +\{\T^b,\T^d,\T_e\} \T^a\T_c + \{\T^a,\T^d,\T_e\} \T_c\T^b + \{\T_c,\T^d,\T_e\}\T^a\T^b\)
\eea
That is, we have only $4$ terms and this expression does not treat the indices $a,b,c,d,e$ on the same footing, and it is not hard to check that the product is not associative. To this end we need to obtain a symmetric expression that at linear order in $\hbar$ involves precisely $10$ terms, which contains all the independent permutations of $\{T^a,\T^b,\T_c\}\T^d\T_e$. The number of such independent permutations equals the number of ways that we can select $2$ indices out of $5$ where the ordering does not matter. 

We can generate the missing $6$ terms by extending the definition of the star-three-product as
\bea
(\T^a * \T_c * \T^b)(\sigma) &=& \lim_{\sigma'\rightarrow \sigma}  \lim_{\sigma''\rightarrow \sigma'} \cr
&&\exp\(\frac{\hbar}{2}\(\{\bullet,\bullet'',\bullet'\} +  \{\bullet'',\bullet',\bullet^{out}\} - \{\bullet,\bullet',\bullet^{out}\} - \{\bullet,\bullet'',\bullet^{out}\}\)\) \cr
&& \T^a(\sigma)\T_c(\sigma') \T^b(\sigma'').
\eea
To describe the meaning of $\bullet^{out}$ we need some more terminology. We are interested in nested products such as $(((\T^a * \T_c * \T^b) * \T_e * \T^d) * \T_g * \T^f$. We will refer to this three-product as a `final' three-product if it is not three-multiplied by anything else. We will refer to $\T^a * \T_c * \T^b$ and $((\T^a * \T_c * \T^b) * \T_e * \T^d)$ as `inner' three-products as they are three-multiplied by further elements. We also want a better name for `further elements'. We will refer to these as outer three-products with respect to a given inner three-product. So for instance with respect to $\T^a * \T_c * \T^b$, both $((\T^a * \T_c * \T^b) * \T_e * \T^d)$ and $(((\T^a * \T_c * \T^b) * \T_e * \T^d) * \T_g * \T^f$ will be referred to as `outer three-products'. Hence an outer three-product can be either an inner or the final three-product. With this terminology we can now explain how the $\bullet^{out}$ acts. If the three-product is final, then $\bullet^{out} = 0$. In other words we compute the final three-product by using the naive definition in Eq (\ref{naive}). The difference from Eq (\ref{naive}) arises when we compute the inner three-products where $\bullet^{out}$ will act on all outer three-products after the first limiting procedure has been carried out so that one has identified all the outer coordinates and wecan evaluate the derivative at the single point $\sigma^{out}$. To be concrete, if we have one outer star-product, then we start out with five different points, or bullets, $\bullet,\bullet',\bullet''$ for the inner three-product and two more, say $\bullet''',\bullet''''$ for the outer and final three-product. Then the limiting procedure for computing the inner three-product is 
\bea
\lim_{\sigma'\rightarrow \sigma} \lim_{\sigma''\rightarrow \sigma'} 
\eea
resulting in an expression at a single point $\sigma$, but where we still have a third bullet $\sigma^{out}$ that acts like a differential operator on the outer three-prouct after we have carried out the first of the two remaining limiting procedures
\bea
\lim_{\sigma''''\rightarrow \sigma'''}
\eea
After this limiting procedure, we identify $\sigma^{out} = \sigma'''$ and we can let $\bullet^{out}$ act from the inner three-product on the outer. Finally we take the final limit
\bea
\lim_{\sigma^{out}\rightarrow \sigma}
\eea
that brings the final three-bracket down to a scalar function defined at the single point $\sigma$. 

We then extend this procedure of computing nested three-products iteratively so that we can compute nested three-products of arbitrary length. 

In Appendix $B$ we show that the star-three-product is associative for arbitrary length of nested three-star-products. Associativity is crucial to everything we do in this paper. But since our proof is very technical, it is put as an Appendix $B$ to this paper.

In Appendix $A$ we demonstrate that the final three-bracket
\ben
[\T^a,\T^b;\T^c] &=& \T^a * \T_c * \T^b - \T^b * \T_c * \T^a\label{braket}
\een
is totally antisymmetric, in the sense that 
\bea
[\T^a,\T^b;\T^c] &=& -[\T^a,\T_c;\T_b].
\eea
It is manifestly antisymmetric in $\T^a,\T^b$. This total antisymmetry has the far-reaching consequence that we can formulate ABJM theory in the manifestly $SO(8)$ invariant form of BLG theory.

\subsection{The three-algebra}
For simplicity let us assume that the star-three-product is defined on a rectilinear three-torus with radii $R_1$, $R_2$, $R_3$, being parametrized by three coordinates $\sigma_{\alpha}\in [0,2\pi]$ where $\alpha = 1,2,3$. We choose the three-algebra generators as
\bea
\T^m &=& e^{im_{\alpha}\sigma^{\alpha}}.
\eea
From our star-three-product we then calculate the three-bracket
\bea
[\T^m,\T^n;\T^p] &=& \T^{m+n-p} \Big(e^{\frac{i\hbar}{2}\{m,n,p\}} e^{\frac{\hbar}{2}\(\{n,p,\bullet\}-\{m,p,\bullet\}+\{m,n,\bullet\}\)}\cr
&& - e^{-\frac{i\hbar}{2}\{m,n,p\}} e^{-\frac{\hbar}{2}\(\{n,p,\bullet\}-\{m,p,\bullet\}+\{m,n,\bullet\}\)}\Big).
\eea
We will think of this as being a three-algebra of the form
\bea
[\T^m,\T^n;\T^p] &=& T^q \hat{f}^{mn}{}_{pq}
\eea 
where $\hat{f}^{mn}{}_{pq}$ contains differential operators. We have substituted some of the differential operators with the corresponding eigenvalues when acting on the corresponding three-algebra generator, but not all differential operators can be eliminated this way. To eliminate them all we act with the hatted structure constants on a constant function which is equal to one everywhere, which we denote as $'1'$,
\bea
f^{mn}{}_{pq} &=& \hat{f}^{mn}{}_{pq} \cdot 1.
\eea
The $f^{mn}{}_{pq}$ are then just numbers, but these do not satisfy the fundamental identity in any sense. They are just some numbers, which in the physical theory correspond to coupling constants. 

We define the inner product as
\ben
\<\X,\Y\> &=& \int \frac{d^3\sigma}{(2\pi)^3} (\X \cdot 1)(\Y \cdot 1).\label{innerproduct}
\een
The inner product can be used to write down a {\sl{gauge invariant}} Lagrangian, despite the $f^{mn}{}_{pq}$ do not satisfy the fundamental identity. If for example we denote eight scalar fields as $X^I = X^I_m \T^m$ then we have a gauge invariant term
\bea
\<[X^I,X^J;X^J],[X^I,X^J;X^K]\> &=& X^I_m X^J_n X^{Kp} f^{mn}{}_{pq} X^{Im'} X^{Jn'} X^{K}_{p'} f^{p'q}{}_{m'n'}.
\eea
Under a gauge transformation, a scalar field $X^I = X^I_m \T^m$ changes infinitesimally by 
\bea
\delta X^I &=& [X^I,\T^m;\T^n]\Lambda^n{}_m
\eea
so the variation $\delta X^I$ is a differential operator. Then $X+\delta X$ is a number (or a function) plus a differential operator. In general we define a scalar field such that it is in the same gauge orbit as $X_m \T^m$. In other words we may always find one representative in each gauge orbit which is a number. This number can indeed be easily extracted. It will be always given by
\bea
X^I_m \T^m &=& X^I \cdot 1.
\eea
This we see directly from $(X + \delta X) \cdot 1 = X$ if $X = X\cdot 1$, and this can be iterated to get a finite variation. The sextic potential is gauge invariant by the fundamental identity which holds for nested three-brackets. The differential operators are needed to have a gauge invariant theory, but this does not appear explicitly in the Lagrangian. 

The Lagrangian is of ABJM form, but expressed in terms of hermitian three-algebra \cite{Bagger:2008se}. We may use the three-algebra we have introduced above. Even though its structure constants are differential operators which act in a particular way on an internal three-torus, it is nevertheless a hermitian three-algebra which obeys the fundamental identity and also the trace-invariance condition. Trace-invariance can be checked in a similar way that one checks trace-invariance of the Nambu-bracket. Now these properties are precisely what we need to write down an ABJM Lagrangian. However this will give ABJM theories with new gauge groups that have not been studied before. The three-brackets that enter the inner-products, and hence appear in the Lagrangian, are always 'final', that is, they are on the form $[X^I,X^J;X^K]\cdot 1$. As we demonstrate in the Appendix A, final brackets are totally antisymmetric in the same fashion as the usual Nambu-bracket. This means that we can use $SO(8)$ triality and map the ABJM Lagrangian into a manifestly $SO(8)$ invariant form of BLG theory. We may need to clarify what we mean by the three-bracket $[\T^m,\T^n;\T^p]$ being totally antisymmetric. First we observe that 
\bea
[\T^m,\T^n;\T^p] \cdot 1 &=& \hbar \{\T^m,\T^n,\T_p\} + \O(\hbar^2).
\eea
Neglecting $\O(\hbar^2)$ this is nothing but the totally antisymmetric Nambu-bracket. What we show in the Appendix A is that this antisymmetry holds to all orders in $\hbar$. The antisymmetry property is transparent when expressed in terms of the Nambu bracket,
\bea
\{\T^m,\T^n,\T_p\} &=& -\{\T^m,\T_p,\T^n\}
\eea
but is a bit more subtle if expressed in terms of the three-bracket, where it reads
\bea
[\T^m,\T^n;\T^p] &=& -[\T^m,\T_p;\T_n].
\eea
In particular the antisymmetry property has a meaning only once we enlarge the three-algebra to include the generators $\T_m$. In the case of a three-torus we have $\T_m = \T^{-m}$ which are again elements in the three-algebra. 

As we will see below in the discussion of fuzzy-tori, it is for the particular values $\hbar = 4\pi R_1R_2R_3/ N$ consistent to truncate the generators to the finite set $m_{\alpha} = 0,...,N-1$ for each $\alpha = 1,2,3$. The three-algebra is then finite-dimensional with $N^3$ generators, but has structure constants which are differential operators.

One may wonder if there is also a conventional way to view the same three-algebra, in which the structure constants are just ordinary numbers. To linear order in $\hbar$ we have 
\bea
[\T^m,\T^n;\T^p] &=& i\hbar \{m,n,p\} \T^{m+n-p} \cr
&& + \hbar \T^{m+n-p}  \(\{n,p,\bullet\}-\{m,p,\bullet\}+\{m,n,\bullet\}\) \cr
&&+ \O(\hbar^2).
\eea
and we may view this as a three-algebra with structure constants which are complex numbers rather than differential operators, at the price of introducing further generators on the form $\T^m \{n,p,\bullet\}$. Let us collectively denote these generators as $\N^A$. The three-algebra then has the structure
\bea
[\T^a,\T^b;\T^c] &=& f^{ab}{}_{cd} \T^d + f^{ab}{}_{cA} \N^A,\cr
[\T^a,\T^b;\N^A] &=& f^{ab}{}_{Ac} \T^c + f^{ab}{}_{AB} \N^B,\cr
[\N^A,\T^b;\T^c] &=& f^{Ab}{}_{cd} \T^d + f^{Ab}{}_{cB} \N^B,\cr
[\N^A,\N^B;\T^c] &=& f^{AB}{}_{cd} \T^d + f^{AB}{}_{cC} \N^C,\cr
[\N^A,\T^b;\N^C] &=& f^{Ab}{}_{Cd} \T^d + f^{Ab}{}_{CD} \N^D,\cr
[\N^A,\N^B;\N^C] &=& f^{AB}{}_{Cd} \T^d + f^{AB}{}_{CD} \N^D.
\eea
Explicitly we may compute brackets which involve $\N^A$ as follows,
\bea
[\N^A,\T^u;\T^{-v}] &=& [[\T^m,\T^n;\T^{-p}],\T^u;\T^{-v}] - [[\T^m,\T^n;\T^{-p}]\cdot 1,\T^u;\T^{-v}]
\eea
We first compute the nested three-bracket
\bea
&&[[\T^m,\T^n;\T^{-p}],\T^u;\T^{-v}]\cdot 1 = 2 \T^{m+n+p+u+v}\cr
&& \times \Bigg(\cos \frac{\hbar}{2}\Big(\{mpn\}+\{mvu\}+\{nvu\}+\{pvu\}\cr
&&+\{pnu\}+\{pnv\}+\{mpu\}+\{mpv\}+\{mnu\}+\{mnv\}\Big)\cr
&& - \cos \frac{\hbar}{2}\Big(\{mpn\}-\{mvu\}-\{nvu\}-\{pvu\}\cr
&&+\{pnu\}+\{pnv\}+\{mpu\}+\{mpv\}+\{mnu\}+\{mnv\}\Big)\Bigg).
\eea
and then the naive three-bracket obtained by neglecting differential operator parts,
\bea
&&[[\T^m,\T^n;\T^{-p}]\cdot 1,\T^u;\T^{-v}]\cdot 1 = 2 \T^{m+n+p+u+v}\cr
&& \times \Bigg(\cos \frac{\hbar}{2}\Big(\{mpn\}+\{mvu\}+\{nvu\}+\{pvu\}\Big)\cr
&& - \cos \frac{\hbar}{2}\Big(\{mpn\}-\{mvu\}-\{nvu\}-\{pvu\}\Big)\Bigg).
\eea
From such a computatation we deduce that
\bea
[\N^A,\T^u;\T^{-v}]\cdot 1 = \O(\hbar^2)
\eea
but is non-zero. The inner product defined as in (\ref{innerproduct}) gives
\bea
\<\T^m,\T^n\> &=& \delta^m_n,\cr
\<\T^m,\N^A\> &=& 0,\cr
\<\N^A,\N^B\> &=& 0.
\eea
Trace-invariance would imply that $f^{A\bullet}{}_{\bullet\bullet} = f^{\bullet\bullet}{}_{A\bullet} = 0$, but the explicit computation above showed that this is true only up to linear order in $\hbar$. At higher orders we have $f^{Am}{}_{np} = \O(\hbar^2)$ and non-vanishing. The only trace-invariance condition that survives to all orders in $\hbar$ is  
\bea
\<[\X,\T^m;\T^n],\Y\> - \<\X,[\Y,\T^n;\T^m]\> &=& 0.
\eea
when $\X$ and $\Y$ are given either by $\X_m \T^m$ and $\Y_m \T^m$ or are gauge variations of these in the form $\X + \delta X$ where $\delta X = [X,\T^m;\T^n]\Lambda^n{}_m$. These gauge variations can also be iterated arbitrarily many times to produce a finite gauge variation. 

Even though we may think on the three-algebra itself as being generated by a finite set $\{\T^m\}$ and an infinite set $\{\N^A\}$ and which has as structure constants usual numbers which are subject to the fundamental identity, we only have trace-invariance among the $\T^m$ generators. A better interpretation of this three-algebra appears to be that the $\T^m$ generate the full three-algebra while the structure constants are differential operators. Being differential operators, it appears that the fundamental identity can not be expressed in terms of $\hat{f}^{mn}{}_{pq}$ alone since these act on the generators $\T^m$ themselves.

\subsection{Dimensional reduction}
As the three-star-product is quite subtle it might be helpful to see an example where we can think on it in terms of usual star-product applied twice. Associativity will then be transparent as the star-product is associative. 

To reduce, we assume that we have circle fiber over a two-manifold. There are two ways of reducing. One is to shrink the size of the circle. Another is to orbifold the circle by a cyclic group. We will consider the latter alternative. We start by writing 
\bea
\T^a * \T_c * \T^b &=& \(\exp \frac{i\hbar}{2} \{k,k',k''\}\) \T^a_k \T_c^{k''} \T^b_{k'} e^{i(k+k'-k'')\sigma}
\eea
in Fourier space. Here
\bea
\{k,k',k''\} &\equiv & \sqrt{g} \epsilon^{\alpha\beta\gamma} k_{\alpha}k'_{\beta}k''_{\gamma}.
\eea
and $g$ is the determinant of the metric on the three-manifold. We note that any sector with a fixed third momentum $k_3$ is closed under the three-bracket since if $k_3 = k'_3 = k''_3$ then the three-bracket generates momenta which have its third component given by $k_3+k'_3-k''_3 = k_3$. We assume the fiber is parametrized by 
\bea
\sigma^3 \in [0,2\pi]
\eea
in which case we can consider a truncation to generators $\T^a_k = \T^a_{k_1 k_2 1}$. That is we restrict to the sector with 
\bea
k_3 &=& 1.
\eea
If we define
\bea
\{k,k'\} &\equiv & \sqrt{g}\epsilon^{\alpha\beta 3} k_{\alpha}k'_{\beta}
\eea
then we have (in this restricted sector)
\ben
\{k,k',k''\} &=& \{k,k'\} + \{k',k''\} + \{k'',k\}.\label{expanding}
\een
This implies that in this sector we can obtain the star-three-product by computing two consecutive star-products that we define as
\bea
\T^a * \T_c &=& \(\exp \frac{i\hbar}{2} \{\bullet,\bullet'\}\) \T^a(\sigma) \T_c(\sigma')
\eea
It is then easy to see that
\bea
(\T^a * \T_c) * \T^b &=& \T^a * \T_c * \T^b
\eea
where in the right-hand side we have the star-three-product. 

It is not obvious that this property would persist when we consider nested star-three-products. However both the star-product and the star-three-product are associative. Associativity is a strong requirement and one could therefore expect that consecutive star-three-products can also be expressed in terms of star-products. 

We will not attempt to prove this in general. Instead we just consider two nested star-three-products. Let us consider 
\bea
(\T^a * \T_c * \T^b) * \T_e * \T^f &=& \exp \frac{i\hbar}{2} \Big( \{k_a,k_b,-k_c\} + \{k_b,-k_c,-k_e+k_d\} - \{k_a,-k_c,-k_e++k_d\} \cr
&& - \{k_a,k_b,-k_e+k_d\}\Big)\cr
&& \exp \frac{i\hbar}{2}\{k_a+k_b-k_c,k_d,-k_e\} \cr
&& \T^a(k_a) \T_c(k_c) \T^b(k_b) \T_e(k_e) \T^d(k_d) e^{i(k_a - k_c + k_b - k_e + k_d)x}.
\eea
We expand out the Nambu brackets according to (\ref{expanding}). We then get, after some cancelations,
\bea
&&\exp \frac{i\hbar}{2}\Big(-\{k_a,k_b\} - \{k_b,k_c\} - \{k_c,k_a\} - \{k_e,k_a\} + \{k_d,k_a\} \cr
&&+ \{k_e,k_d\} + \{k_b,k_e\} + \{k_d,k_b\} - \{k_c,k_e\} - \{k_d,k_c\}\Big).
\eea
If we expand out star-products in two dimensions of the corresponding expression we get
\bea
(((\T^a * \T_c) * \T^b) * \T_f) * \T^e &=& \exp \frac{i\hbar}{2} \Big(\{k_a,-k_c\} + \{k_a-k_c,k_b\} \cr
&& + \{k_a-k_c+k_b,-k_e\} + \{k_a - k_c + k_b - k_e,k_d\}\Big).
\eea
We can now see that the two results agree.

\subsection{Motivating the dimensional reduction}
We would now like to motivate why we should we take $k_3 = 1$ to reduce the star-three-product to a star-product. Lending an idea from \cite{Kim:2008gn}, we postulate that in order to dimensionally reduce, we shall orbifold the three-manifold on which the star-three-product is originally defined. Let us take our three-manifold to be a three-torus and impose the orbifold identification
\bea
\T^a &\sim& e^{\frac{2\pi i}{K}} \T^a.
\eea
We parametrize the three-torus by $\sigma^{\alpha}$ for $\alpha = 1,2,3$. Each of which runs over the interval $[0,2\pi]$. When we orbifold the torus, we shall also impose the identification
\bea
\sigma^3 &\sim & \sigma^3 + \frac{2\pi}{K}.
\eea
This orbifolding prescription is speculative. 

As we will see in more detail below, we can consider finite-dimensional truncations of the three-algebra generators. For finite $K$ we can truncate the three-algebra to functions (We use as three-algebra index $\vec{m}$ in place of $a$)
\bea
\T^{\vec{m}} &=& e^{i(m_1 \sigma^1 + m_2 \sigma^2 + m_3 \sigma^3)}
\eea
on the orbifolded three-torus where 
\bea
m_1,m_2 &=& 1,...,N,\cr
m_3 &=& 1, 1+K, ..., 1+\[\frac{N-1}{K}\]K.
\eea
Here $N$ is a cut-off on the frequencies, and this is a consistent finite-dimensional truncation of three-algebra generators which is such that these generators close among themselves under the three-bracket. 

Dimensional reduction should mean that one direction shrinks to zero size. Our prescription for dimensional reduction amounts to taking the limit of a large $K$. By increasing $K$ the size of the $\sigma^3$ direction shrinks. When $K$ exceeds that value $N$ the only possible value for $m_3$ is
\bea
m_3 &=& 1.
\eea
In our previous subsection we wrote $k_3$ in place of $m_3$.

\subsection{Three-algebra homomorphism}
We have seen that we can truncate the three-algebra generated by functions 
\bea
\T^m &=& e^{im_{\alpha}\sigma^{\alpha}}
\eea
on the orbifolded three-torus, to the sector
\bea
m_3 &=& 1.
\eea
If we define
\bea
\{m,n,p\} &\equiv & \sqrt{g} \epsilon^{\alpha\beta\gamma} m_{\alpha}n_{\beta}p_{\gamma},\cr
\{m,n\} &\equiv & \sqrt{g}\epsilon^{\alpha\beta 3} m_{\alpha}n_{\beta}.
\eea
then we have
\bea
\{m,n,p\} &=& \{m,n\} + \{n,p\} + \{p,m\}.
\eea
Once we make this truncation, the star-three-product
\bea
\T^m * \T_p * \T^n &=& e^{\frac{i\hbar}{2}\{m,n,p\}} \T^{m+n-p}
\eea
can be viewed as two consecutive star-products, 
\bea
\T^m * \T_p &=& e^{\frac{i\hbar}{2}\{m,-p\}} \T^{m-p},\cr
\T^{m-p} * \T^n &=& e^{\frac{i\hbar}{2}\{m-p,n\}} \T^{m+n-p}.
\eea
We can now compute things in two different ways. Either we use the two dimensional star product (two-torus language), or we use the star-three-product and restrict ourselves to the sector $m_3 = 1$ (three-torus language). Both ways give the same answer. The functions $\T^{m} = \T^{m_1m_2m_3}$ generate three-algebras which have no matrix realizations. If we truncate to functions $\T^m = \T^{m_1m_2 1}$, then it is easy to see that these generate a sub-three-algebra. These are related to a two-torus and can therefore be related to matrices associated with a fuzzy two-torus. To find the appropriate functions which we can relate to matrices, we first define the basic functions 
\bea
\U &=& e^{i\sigma^1},\cr
\V &=& e^{i\sigma^2}
\eea
These are subject to 
\bea
\U * \V &=& e^{\frac{i\hbar}{\sqrt{g}}} \V * \U,\cr
\U^* &=& \U^{-1},\cr
\V^* &=& \V^{-1}
\eea
Given a star-product it is natural to define
\bea
t^{m_1 m_2} &=& \U^{m_1} * \V^{m_2}
\eea
where all products are star products. Notice that this makes sense only if we truncate to $m_3 = 1$ or we consider the two-torus where we have a star product which can be used to multiply two functions. We now get
\bea
t^{m_1 m_2} &=& e^{\frac{i\hbar}{2}\{(m_1 0),(0 m_2)\}} e^{i(m_1 \sigma^1 + m_2 \sigma^2)}\cr
&=& e^{\frac{i\hbar}{2\sqrt{g}}m_1 m_2} e^{i(m_1 \sigma^1 + m_2 \sigma^2)}.
\eea
That is we define $t^m$ with an additional phase factor. Without this phase factor, that is for $\T^m = e^{i(m_1\sigma^1+m_2\sigma^2+\sigma^3)}$ in the three-torus language say, we get
\bea
[\T^m,\T^n;\T^p] &=& \(e^{\frac{i\hbar}{2}\{m,n,p\}} - e^{-\frac{i\hbar}{2}\{m,n,p\}}\)\T^{m+n-p}.
\eea
Including the phase factor we instead get
\bea
[t^m,t^n;t^p] &=& e^{\frac{i\hbar}{2\sqrt{g}}\(m_1m_2 + n_1n_2 - p_1p_2 - (m+n-p)_1 (m+n-p)_2\)}\cr
&&\times \(e^{\frac{i\hbar}{2}\{m,n,p\}} - e^{-\frac{i\hbar}{2}\{m,n,p\}}\)t^{m+n-p}.\label{threealg}
\eea

We now turn to matrices. We define
\bea
T^{m_1 m_2} &=& U^{m_1} V^{m_2}
\eea
where $U$ and $V$ are basic matrices subject to 
\bea
UV &=& e^{\frac{2\pi i}{N}} VU,\cr
U^{\dag} &=& U^{-1},\cr
V^{\dag} &=& V^{-1},\cr
U^N &=& 1,\cr
V^N &=& 1
\eea
and all multiplications are given by matrix multiplications. Explicitly we can take them as
\bea
U &=& \(\begin{array}{ccccc}
0 & 1 & 0 &     & 0\\
0 & 0 & 1 &     & 0\\
  &   & \ddots  &  \\
1 & 0 & 0  &     & 0
\end{array}\),\cr
V &=& \(\begin{array}{cccc}
1 & 0 &  & 0\\
0 & e^{i\t \hbar} &  & 0\\
  &            & \ddots  &  \\
0 & 0 &  & e^{i(N-1)\t \hbar}
\end{array}\)
\eea
where $\t \hbar=2\pi/N$.

The map
\bea
t^{m_1 m_2} \rightarrow T^{m_1 m_2}
\eea
is a three-algebra homomorphism. To show this we first note that
\bea
U^{m_1} V^{m_2} &=& e^{\frac{2\pi m_1 m_2 i}{N}} V^{m_2} U^{m_1}
\eea
Then we compute
\bea
[T^m,T^n;T^p] &=& U^{m_1} V^{m_2} \(U^{p_1} V^{p_2}\)^{\dag} U^{n_1} V^{n_2} - U^{n_1} V^{n_2} \(U^{p_1} V^{p_2}\)^{\dag} U^{m_1} V^{m_2}\cr
&=& \(e^{-\frac{2\pi i(m_2-p_2)(m_1-p_1)}{N}} - e^{-\frac{2\pi i(n_2-p_2)(m_1-p_1)}{N}}\) U^{m_1+n_1-p_1}V^{m_2+n_2-p_2}
\eea
We can write this in a neater form as
\bea
[T^m,T^n;T^p] &=& e^{-\frac{\pi i}{N}\((m,n) - (p,n) - (m,p) - 2(p,p)\)}\cr
&&\times \(e^{\frac{\pi i}{N}\([m,n]+[n,p]+[p,m]\)} - e^{-\frac{\pi i}{N}\([m,n]+[n,p]+[p,m]\)}\)T^{m+n-p}
\eea
where 
\bea
[m,n] &\equiv & m_1 n_2 - m_2 n_1,\cr
(m,n) &\equiv & m_1 n_2 + m_2 n_1.
\eea
The three-algebra we get from the matrix calculation agrees with the three-algebra we get using star products of functions, if we take
\bea
\frac{\hbar}{\sqrt{g}} &=& \frac{2\pi}{N}.
\eea

On the one hand the matrices generate $U(N)$ Lie algebra. On the other hand they generate the three-algebra of unitary $N\times N$ matrices. 

Unit normalized trace forms are given by
\bea
\<\t \T^{\vec{m}},\t \T^{\vec{n}}\> &=& \int \frac{d^2\sigma}{(2\pi)^2}\t \T^{\vec{m}} \t \T_{\vec{n}}
\eea
and 
\bea
\<T^{\vec{m}},T^{\vec{n}}\> &=& \frac{1}{N} \tr\(T^{\vec{m}} T_{\vec{n}}\)
\eea
respectively.

\subsection{Fuzzy tori}
So far we have thought about matrices $U$ and $V$ as generators of a three-algebra. The more familiar interpretation is that they describe a fuzzy two-torus \cite{Fairlie:1989vv}, \cite{Madore:1991bw}, \cite{Bigatti:2001gy}. These matrices can be mapped into functions $e^{i\sigma^1}$ and $e^{i\sigma^2}$ that live on the two-torus. The fuzziness comes about from the fact that one may realize the algebra $UV = e^{\frac{2\pi i}{N}} VU$ by demanding that the coordinates do not quite commute,
\bea
[\sigma^1,\sigma^2] &=& i \t\hbar.
\eea
The non-commutativity parameter $\t \hbar$ can be derived by using the BCH formula 
\bea
e^{i\sigma^1} e^{i\sigma^2} &=& e^{i\sigma^1 + i\sigma^2 + \frac{1}{2}[i\sigma^1,i\sigma^2]}
\eea
which is an exact formula when $[\sigma^1,\sigma^2]$ is a c-number. We then get
\bea
\t \hbar &=& \frac{2\pi}{N}.
\eea
This goes to zero as $N$ goes to infinity which means that we get a classical torus in this limit. A concrete realization is in terms of functions and the star-product. The commutator is given by
\bea
[\sigma^1,\sigma^2] = \sigma^1 * \sigma^2 - \sigma^2 * \sigma^1 = i\hbar \{\sigma^1,\sigma^2\}.
\eea
In our convention we have $\{\sigma^1,\sigma^2\} = 1/\sqrt{g_2}$ where $g_2$ denotes the determinant of the metric on the two-torus, so that $\t \hbar = \hbar/\sqrt{g_2}$. We have Lie algebra generators 
\bea
\T^{m} &=& e^{\frac{i\t\hbar}{2}m_1m_2} e^{i(m_1\sigma^1+m_2\sigma^2)}
\eea
for $m_1,m_2=0,...,N-1$. These generate $U(N)$. Commutation relations are
\ben
[\T^m,\T^n] &=& \(e^{i\t\hbar m_2n_1}-e^{i\t\hbar m_1 n_2}\) \T^{m+n}\label{Liealgebra}
\een
Here it may seem like one could take any value on $\t \hbar$. However this is so only if we include the infinite set of generators $\T^m$ in the Lie algebra. If we consider a finite truncation, or equivalently impose the identifications
\bea
\T^{m_1m_2} &\sim & \T^{m_1+N,m_2},\cr
\T^{m_1m_2} &\sim & \T^{m_1,m_2+N}
\eea
and represent elements of the algebra as equivalence classes $[\T^m]$, then the Jacobi identity holds only when $e^{i\t\hbar N} = 1$. We see that under that condition the Lie algebra (\ref{Liealgebra}) is invariant under $m_{\alpha} \rightarrow m_{\alpha} + N$. The Jacobi identity holds for the infinite-dimensional Lie algebra by associativity of the star product. The truncation obtained by imposing the periodic identification is a consistent truncation when the structure constants also obey the same periodicity. This is the case for the above values of $\hbar$.

For the fuzzy three-torus we have no realizations in terms of matrices. But we have functions and a star-three-product. That is enough to be able to realize the fuzzy three-torus. A fuzzy three-torus is generated by functions $\U = e^{i\sigma^1}$, $\V = e^{i\sigma^2}$ and $\W = e^{i\sigma^3}$ where $\sigma^{\alpha} \in [0,2\pi]$. The coordinates do not quite three-commute but are subject to
\bea
[\sigma^1,\sigma^2;\sigma^3] &=& \t \hbar.
\eea
Concretely
\bea
[\sigma^1,\sigma^2;\sigma^3] = \sigma^1 * \sigma^3 * \sigma^2 - \sigma^2 * \sigma^3 * \sigma^1 = \hbar \{\sigma^1,\sigma^2,\sigma^3\}
\eea
and we have $\{\sigma^1,\sigma^2,\sigma^3\} = 1/\sqrt{g_3}$ where $g_3$ denotes the determinant of the metric on the three-torus. For suitably chosen values of $\hbar$ we may consider finite truncation of the three-algebra to a finite set of generators 
\bea
\T^{m} &=& \U^{m_1} * \V^{m_2} * \W^{m_3}
\eea
where $m_{\alpha} = 0,...,N-1$ for some finite $N$. We can determine $\hbar$ by requiring the three-algebra be invariant under $m_{\alpha} \rightarrow m_{\alpha} + N$. This implies that $e^{i \t \hbar N/2} = 1$. The fundamental identity is automatic for infinite-dimensional three-algebra by associativity of the star-three-product. The truncation obtained by the periodic identification is a consistent finite truncation when the three-algebra structure constants obey the same periodicity. This is the case for the above values of $\hbar$.

For $N=1$ we have a trivial three-algebra. For $N=2$ we find $2^3 = 8$ different three-algebra generators $\T^{000}, \T^{001}, \T^{010}, ..., \T^{111}$. We are now in a new regime of three-algebras that have not been studied before and which have no matrix realizations and yet they are finite-dimensional.

\section{The M two-brane Lagrangian}
In order to deconstruct the M five-brane and take the continuum limit in a controlled way, and in particular keep track of the numerical value of the coupling constant, we should formulate M two-brane theory in terms of the star-three-product. 

If we realize the three-bracket by functions being multiplied by star-three-products, then as we have seen the three-bracket becomes totally antisymmetric. This means that the Lagrangian can be written in the $SO(8)$ invariant form generalizing the Lagrangian obtained by Bagger and Lambert for real three-algebras \cite{Bagger:2007jr}. Of course there must be a subtlety involved here related with the Chern-Simons level $k$. We do not have $SO(8)$ invariance unless the level is $k=1,2$, if the gauge group is $U(N_1)\times U(N_2)$, and here we only consider cases where $N_1=N_2=N$. We must thus break $SO(8)$ for $k>2$. There is indeed a natural way of breaking $SO(8)$ if we formulate ABJM theory in terms of star-three-product on an auxiliary space. Now if the auxiliary space means that three directions are compactified on a three-torus, then this is extremely subtle as that space does not possess much rotational symmetry, so $SO(8)$ is not there to start with not even if $k=1$. We may consider a three-sphere embedded in $\mb{C}^4$ instead. We can break $SO(8)$ by orbifolding and consider instead ${\mb{C}}^4/{\mb{Z}}_k$ where the ${\mb{Z}}_k$ acts like $Z^A \rightarrow e^{\frac{2\pi i}{k}} Z^A$ if $Z^A$ denote the four complex coordinates in ${\mb{C}}^4$. This orbifolding induces a natural orbifolding on the three-sphere embedded in ${\mb{C}}^4/{\mb{Z}}_k$. If we view the three-sphere as a Hopf fibration, then orbifolding acts on the fiber direction so as to get $S^1/{\mb{Z}_k}$. Using star-three-product on this orbifolded three-sphere we get an M two-brane action with $SO(8)$ broken explicitly down to $SU(4)\times U(1)$ for $k>2$. With this prescription for how to associated an auxiliary space to with Chern-Simons level $k$, the issue regarding supersymmetry enhancement for level $k=1,2$ is completely eliminated. However all is not solved by this. We still have to understand why we should use this ${\mb{Z}_k}$-orbifold and why the $k$ in the orbifold should be at all related with the $k$ in the Chern-Simons level. These questions we will not answer in this paper. An intuitive picture is that in the large $N$ limit, matrices start to commute and we find a classical sphere. But commuting matrices resemble $U(1)\times U(1)$ gauge group. The orbifold ${\mb{C}}^4/{\mb{Z}}_k$ has been derived from $U(1)\times U(1)$ ABJM theory at level $k$. Perhaps a similar simplication occurs also for $U(\infty)\times U(\infty)$ gauge group.

We expect that one can derive our manifestly $SO(8)$ invariant M two-brane Lagrangian on  target space ${\mb{C}}^4/{\mb{Z}}_k$ from ABJM theory in a similar way as the manifestly invariant $SO(8)$ invariant sigma model was derived on ${\mb{C}}^4/{\mb{Z}}_k$ from $U(1)\times U(1)$ ABJM theory. In both cases $SO(8)$ is manifest in the Lagrangian, but is expliclity broken by the orbifold. 

A natural guess of how this extends to the case when target space is compactified on a torus would be that we should consider the orbifold ${\mb{C}}^4/{\mb{Z}_k}/\sim$ where $\sim$ is the torus identification $Z^A \sim Z^A + 2\pi R^A$. However it is not clear which points in $\mb{C}^4$ really gets identified as $\mb{Z}_k$ and $\sim$ act so differently on each point (one is a translation, the other is a rotation), and so we will in this paper only assume a three-torus with $k=1$.  

We will work with the M two-brane Lagrangian expressed as
\bea
\L &=& \L_m + \L_{CS}
\eea
plus the supersymmetric completion which we do not write out. Here
\bea
\L_{m} &=& -\frac{1}{2}\<D_{\mu}X^I,D^{\mu}X^I\> - \frac{\lambda^2}{12} \<[X^I,X^J;X^K],[X^I,X^J;X^K]\>,
\eea
\bea
\L_{CS} &=& \frac{\lambda}{2}\<[\T^a,\T^b;\T^c],\T^d\>A^c{}_b dA^d{}_a - \frac{\lambda^2}{3} \<[\T^a,\T^c;\T^d],[\T^b,\T^f;\T^e]\>A^b{}_a A^d{}_c A^f{}_e
\eea
The covariant derivative is given by
\bea
D_{\mu} X^I &=& \partial_{\mu} X^I + \lambda[X^I,\T^a;\T^b]A_{\mu}{}^b{}_a.
\eea
Gauge variations are
\ben
\delta X^I &=& -\lambda[X^I,\T^a;\T^b]\Lambda^b{}_a,\label{etta},\\
\delta A_{\mu} &=& D_{\mu}\Lambda.\label{tvaa}
\een
Explicitly the second variation reads
\bea
[\bullet,\T^a;\T^b]\delta A_{\mu}{}^b{}_a &=& [\bullet,\T^a;\T^b]\partial_{\mu}\Lambda^b{}_a \cr
&&+ \lambda\([[\bullet,\T^c;\T^d],\T^a;\T^b]-[[\bullet,\T^a;\T^b],\T^c;\T^d]\)\Lambda^d{}_c A_{\mu}{}^b{}_a.
\eea
If we choose real three-algebra generators, $\T^a = \T_a$ we get the Bagger-Lambert Lagrangian with trace-form $h_{ab} = \delta_{ab}$. Here $\lambda$ is a coupling constant that could have been absorbed into the definition of the three-bracket. Though we will not do that, but will write $\lambda$ explicitly, and we will always define the three-bracket as $[\T^a,\T^b;\T^c] = \T^a*\T_c*T^b - \T^b*\T_c*\T^a$. 

As said, this form of the Lagrangian is valid only if the three-bracket is totally antisymmetric. This is not the case if we realize it by matrices. If we would, then we would have to use the ABJM Lagrangian. Here we will realize the three-algebra by functions 
\bea
\T^m &=& e^{im_{\alpha}\sigma^{\alpha}}
\eea
on a three-torus, in which case the three-bracket is totally antisymmetric and we can keep manifest $SO(8)$ symmetry in the Lagrangian. The unit normalized inner product is given by 
\bea
\<\T^m,\T^n\> &=& \int \frac{d^3\sigma}{(2\pi)^3} \T^m \T_n.
\eea

Let us consider the Lagrangian to linear order in $\hbar$. Here we have
\bea
[X^I,X^J;X^K] &=& \hbar \{X^I,X^J,X^K\} + \O(\hbar^2).
\eea
In the Lagrangian we then have $\lambda\hbar$ multiplying the Nambu bracket. We can get rid of this factor by rescaling the fields 
\bea
X^I &\rightarrow & \frac{1}{\sqrt{\lambda \hbar}} X^I,\cr
A &\rightarrow & \frac{1}{\lambda \hbar} A.
\eea
This brings the factor $\frac{1}{\lambda\hbar}$ in front of the whole Nambu Lagrangian instead. 

We can not relate the three-torus to matrices. To this end we must truncate to the two-torus by taking $m_3 = 1$. We then get
\bea
\hbar &=& \frac{2\pi}{N}\sqrt{g}.
\eea
The unit normalized norm of matrices $T^m = U^{m_1} V^{m_2}$ is given by 
\bea
\<T^m,T^n\> &=& \frac{1}{N} \tr (T^m T_n)
\eea
where $T_m$ denotes the hermitian conjugate matrix of $T^m$. The Lagrangian then becomes
\bea
\L_{m} &=& -\frac{1}{2N}\tr(D_{\mu}X^I D^{\mu}X^I) - \frac{\lambda^2}{12 N} \tr([X^I,X^J;X^K][X^I,X^J;X^K]),
\eea
\bea
\L_{CS} &=& \frac{\lambda}{2N}\tr([T^a,T^b;T^c]T_d) A^c{}_b dA^d{}_a - \frac{\lambda^2}{3N} \tr([T^a,T^c;T^d][T^b,T^f;T^e]^{\dag})A^b{}_a A^d{}_c A^f{}_e.
\eea
If we rescale the fields we can get a factor $\frac{1}{N\lambda}$ in front of the whole Lagrangian, and remove the $\lambda$ (or put it to one) in all interaction terms. In the matrix realization we define adjoint gauge fields 
\bea
A^R_{\mu} &=& T_b T^a A_{\mu}{}^b{}_a,\cr
A^L_{\mu} &=& T^a T_b A_{\mu}{}^b{}_a
\eea
of the gauge group $U(N)_L\times U(N)_R$. The covariant derivative then becomes
\bea
D_{\mu} Z^A &=& \partial_{\mu} Z^A + Z^A A_{\mu}^R - A_{\mu}^L Z^A.
\eea
The Chern-Simons term becomes 
\bea
\L_{CS} &=& \frac{1}{2N\lambda} \tr\(A^R dA^R - \frac{2}{3} (A^R)^3\) - \frac{1}{2N\lambda} \tr\(A^L dA^L - \frac{2}{3} (A^L)^3\).
\eea
We conclude that 
\bea
\frac{1}{2N\lambda} &=& \frac{k}{4\pi}
\eea
where $k$ is the integer quantized Chern-Simons level. 

Putting two things together we have that
\bea
\frac{1}{\lambda\hbar} &=& \frac{N^2 k}{4\pi^2 \sqrt{g}}
\eea
multiplies the whole Lagrangian as
\bea
\L &=& \frac{N^2 k}{4\pi^2 \sqrt{g}} \(\L_m + \L_{CS}\)
\eea
where
\bea
\L_{m} &=& -\frac{1}{2}\<D_{\mu}X^I,D^{\mu}X^I\> - \frac{1}{12} \<\{X^I,X^J,X^K\},\{X^I,X^J,X^K\}\>,
\eea
\bea
\L_{CS} &=& \frac{1}{2}\<\{\T^a,\T^b,\T_c\},\T^d\>A^c{}_b dA^d{}_a - \frac{1}{3} \<\{\T^a,\T^c,\T_d\},\{\T^b,\T^f,\T_e\}\>A^b{}_a A^d{}_c A^f{}_e.
\eea
This form of the Lagrangian is convenient in order to connect the M two-brane with the M five-brane.

\section{Deconstructing single M five-brane}
We will now deconstruct a single M five-brane from multiple M two-branes. As it will turn out, the internal fuzzy three-torus on which the three-algebra generators $\T^a$ are defined, will become part of the M five-brane world-volume. It should be noted that we start out with a three-torus which is purely auxiliary from the M two-brane point of view. For example we normalize the inner product as $\int d^3 \sigma/(2\pi)^3$ and there is no reference to a metric on the three-torus. But in the deconstructed theory we get the reparametrization invariant measure $\int d^3 \sigma \sqrt{g}$ on the three-torus, where $g$ denotes the determinant of the metric. The three-torus which started out its life as an auxiliary space becomes a physical space in the deconstructed theory.

Let us expand the M two-brane Lagrangian about some $x^{\mu}$-independent but $\sigma^{\alpha}$-dependent background scalar field configuration as
\bea
X^I &=& T^I + Y^I
\eea
where $T^I = T^I(\sigma)$ is the background and $Y^I=Y^I(x,\sigma)$ the fluctuation. We will specify a suitable background in more detail later on. Let us decompose the fluctuations into a transverse and a parallel part as
\bea
Y^I &=& Y^{\alpha} \partial_{\alpha} T^I + Y^A \partial_A T^I
\eea
Here $\partial_A = \partial/\partial T^A$ if we split $T^I = (T^{\alpha},T^A)$. We also define
\bea
A_{\mu}{}^b{}_a \T^a \partial_{\alpha} \T_b &=& -\frac{2\pi\hbar}{\sqrt{k}} B_{\mu\alpha}.
\eea
We define the field strength as
\bea
H_{\alpha\beta\gamma} &=& 3\partial_{[\alpha}B_{\beta\gamma]},\cr
H_{\mu\alpha\beta} &=& \partial_{\mu}B_{\alpha\beta} + \partial_{\beta}B_{\mu\alpha} + \partial_{\alpha} B_{\beta\mu}.
\eea

M five-brane dynamics starts at quadratic order in fluctuation fields. It is here that we find all the familiar terms of the M five-brane Lagrangian. We first expand
\bea
D_{\mu}X^I &=& \frac{1}{2\sqrt{k}}\sqrt{g}\epsilon^{\alpha\beta\gamma} H_{\mu\beta\gamma} \partial_{\alpha}T^I + \partial_{\mu}Y^A \partial_A T^I
\eea
and then we compute term by term
\bea
-\frac{1}{2}\<D_{\mu}X^I,D^{\mu}X^I\> &=& -\frac{1}{2}\int \frac{d^3 \sigma}{(2\pi)^3} \(\partial_{\mu}Y^A \partial^{\mu}Y^A + \frac{1}{k}H_{\mu\alpha\beta}H^{\mu\alpha\beta}\),
\eea
\bea
-\frac{1}{2}\<\{X^I,X^J.X^K\},\{X^I,X^J,X^K\}\>|_{quadratic}
&=& -\frac{1}{2}\int \frac{d^3 \sigma}{(2\pi)^3} \(\partial_{\alpha}Y^{\alpha}\partial_{\beta}Y^{\beta} + g^{\alpha\beta}\partial_{\alpha}Y^A \partial_{\beta}Y^A\),
\eea
\bea
\frac{1}{2}\epsilon^{\mu\nu\lambda}A_{\mu}{}^b{}_a \partial_{\nu} A_{\lambda}{}^c{}_d \<T^a,\{T^b,T^c,T_d\}\> &=& -\frac{1}{2k}\int \frac{d^3 \sigma}{(2\pi)^3} \sqrt{g}\epsilon^{\alpha\beta\gamma}\epsilon^{\mu\nu\lambda}  \partial_{\beta} B_{\mu\alpha} \partial_{\nu} B_{\lambda\gamma}
\eea
We next put
\bea
Y^{\alpha} &=& \frac{4\pi^2}{N\sqrt{k}}\frac{g}{2}\epsilon^{\alpha\beta\gamma}B_{\beta\gamma},\cr
Y^A &=& \frac{4\pi^2}{N\sqrt{k}}\sqrt{g}\phi^A.
\eea
We then get the result
\bea
\L &=& -\frac{1}{4\pi} \int d^3 \sigma \sqrt{g} \(\partial_{\mu} \phi^A \partial^{\mu} \phi^A + \partial_{\alpha} \phi^A \partial^{\alpha} \phi^A\)\cr
&& - \frac{1}{4\pi} \int d^3 \sigma \sqrt{g} \(\frac{1}{6} H_{\alpha\beta\gamma} H^{\alpha\beta\gamma} + \frac{1}{2} H_{\mu\alpha\beta} H^{\mu\alpha\beta} + \epsilon^{\mu\nu\lambda}\epsilon^{\alpha\beta\gamma} \partial_{\beta}B_{\mu\alpha} \partial_{\nu}B_{\lambda\gamma}\).
\eea¨
This is now the M five-brane Lagrangian (integrated over three of the six world-volume coordinates) with the value of the coupling constant which is fixed by selfduality of the gauge field strength \cite{Gustavsson:2001wa}, \cite{Henningson:2004dh}. The convention used is such that we have the Dirac charge quantization 
\bea
\int H \in 2\pi Z
\eea
However our logic was to make a choice of numerical constants so that we got this result. 

As one consequence of this choice we get
\bea
\int \frac{d^3 \sigma}{(2\pi)^3} \partial_{\alpha}Y^{\alpha} &=& \frac{1}{2\pi N\sqrt{k}}\int d^3\sigma \frac{g}{6}\epsilon^{\alpha\beta\gamma}H_{\alpha\beta\gamma}.
\eea
If we now restrict to $k=1$ we may interpret this equation as saying that
\bea
\int H &\in & 2\pi N \mb{Z}.
\eea
We do not understand why we get the presence of a factor $N$, but at least this does not violate the general Dirac quantization condition. But why we get a restricted set of allowed charges on a three-torus is not clear to us. Perhaps this is related to that we have M two-branes in a bound state with an M five-brane. To satisfy this condition we may impose the following periodicity conditions on the fluctuation fields, 
\ben
Y^{\alpha}(2\pi) &=& Y^{\alpha}(0) + 2\pi w^{\alpha}\label{choice}
\een
where $w^{\alpha} \in {\mb{Z}}$ are a winding number from (one-cycles of) one three-torus to (one-cycles of) another, via the map $\sigma^{\alpha} \mapsto Y^{\alpha}$. So far we have not specified the background field configuration $T^I$ about which we expand the eight scalar fields. We will now assume that the Chern-Simons level is $k=1$ and take a three-torus as background,
\ben
X^{\alpha} & \sim & X^{\alpha} + 2\pi R^{\alpha}.\label{torusid}
\een
That is, we take 
\ben
T^{\alpha} &=& R^{\alpha} \sigma^{\alpha}.\label{vev}
\een
From the relation
\bea
X^{\alpha} &=& T^{\alpha} + Y^{\alpha}\partial_{\alpha}T^{\alpha}\cr
&=& R^{\alpha}(\sigma^{\alpha} + Y^{\alpha}).
\eea
we see that the choice of boundary conditions (\ref{choice}) is consistent with the torus identification (\ref{torusid}).

For higher values of $k$ we expect some kind of orbifolded three-torus rather than the background we specified above. It now indeed seems extremely subtle how to realize a background which can give us a factor of $\sqrt{k}$. This will be needed in order to not violate the Dirac charge quantization condition $\int H \in 2\pi \mb{Z}$. 

For Abelian M five-brane theory all higher order terms in fluctuation fields will be suppressed by at least one power of $\sqrt{\hbar}$. At this stage we see no trouble in taking the decompactification limits $R_{\alpha}\rightarrow \infty$. The commutative limit $\hbar\rightarrow 0$ and the decompactification limits $R_{\alpha}\rightarrow \infty$ can be taken independently of each other. 

So far we have been ignorant about zeroth and linear order contributions in $\hbar$ to the induced M five-brane Lagrangian. It is time to remedy this. Let us choose the metric on the torus 
\bea
(ds)^2 &=& g_{\alpha\beta} d\sigma^{\alpha}d\sigma^{\beta} 
\eea
such that 
\bea
g_{\alpha\beta} &=& R_{\alpha}^2 \delta_{\alpha\beta}.
\eea
Then the square root determinant is
\bea
\sqrt{g} &=& R_1 R_2 R_3.
\eea
We now expand the Lagrangian about the torus order by order in powers of fluctuation fields. To zeroth order we have (omitting the overall factor of $N^2/(4\pi^2 \sqrt{g})$, to be reinserted later)
\bea
\L &=& -\frac{1}{12}\<\{T^I,T^J,T^K\},\{T^I,T^J,T^K\}\>
\eea
and this formula is exact since all higher order derivatives of the background fields $T^I$ vanish. We compute the Nambu bracket according to 
\bea
\{T^1,T^2,T^3\} &=& \sqrt{g} \epsilon^{\alpha\beta\gamma} \partial_{\alpha}T^1 \partial_{\beta} T^2 \partial_{\gamma} T^3
\eea
We rise indices on the epsilon tensor by the inverse metric and define $\epsilon_{123} = 1$ so
\bea
\{T^1,T^2,T^3\} = \frac{1}{\sqrt{g}} R_1 R_2 R_3 = 1
\eea
We then get the zeroth order result
\bea
\L = -\frac{1}{2} \int \frac{d^3 \sigma}{(2\pi)^3} = -\frac{1}{2}
\eea
which is the exact result (viewed as an $\hbar$ expansion). 

At linear order we have 
\bea
\L &=& -\frac{1}{2}\<[Y^I,T^J,T^K],[T^I,T^J,T^K]\>.
\eea
This we evaluate to
\bea
\L = - \int \frac{d^3\sigma}{(2\pi)^3} \partial_{\alpha} Y^{\alpha} = -\frac{1}{2\pi N} \int H.
\eea

In a fully Lorentz covariant formulation, the Lagrangian in a background $C$ field should be modified as 
\bea
\L_C &=& \frac{1}{8\pi}\(-\frac{1}{2}h\wedge * h + \mu e h\wedge C\)
\eea
where $\mu = \pm 1$ (the sign determines the chirality) and
\bea
h &=& dB + e C
\eea
is the gauge invariant field strength. This construction is made such that $h$ couples only to the (anti)selfdual part $C_{(\mu)} = (C - \mu *C)/2$ \cite{Witten:1996hc}. Let us denote $SO(1,5)$ vector indices as $M,N,... = 0,1,...,5$. Then expand the Lagrangian as
\bea
\L_{C} &=& -\frac{1}{8\pi}\(\frac{1}{12} H_{MNP}H^{MNP} + \frac{\mu e}{6} H_{MNP} C^{MNP}_{(\mu)} + \frac{e^2}{12} C_{MNP} C^{MNP}\).
\eea
even though this does not keep manifest gauge invariance, the purpose of this expression being to match with the result we got from the deconstruction. Let us choose 
\bea
C_{\alpha\beta\gamma} &=& C\epsilon_{\alpha\beta\gamma}
\eea
and all other components to vanish. Then we get
\bea
\L_{C} &=& -\frac{1}{8\pi}\(\frac{1}{12} H_{MNP}H^{MNP} + 2\mu eC \frac{g}{6}\epsilon^{\alpha\beta\gamma} H_{\alpha\beta\gamma}  + \frac{e^2C^2}{2}\).
\eea
modulo one subtlety. Formally we have assumed that $H_{MNP}$ is selfdual when expanding out the second term as $H_{MNP} C^{MNP}_{(\mu)} = 2 H_{\alpha\beta\gamma} C^{\alpha\beta\gamma}$. In the $SO(1,2)\times SO(3)$ formulation the gauge field components $B_{\mu\nu}$ are absent so there would be no term $H_{\mu\nu\lambda} C^{\mu\nu\lambda}_{(\mu)}$. More work will be needed to see that this formalism gives a fully Lorentz covariant quantum theory.

We would now like to match this expected result, with the result that we have got in our deconstruction. Deconstruction gave us a Lagrangian with partially broken Lorentz symmetry. But let us be ignorant about this subtlety here, and let us write the relevant terms of the deconstructed Lagrangian in the Lorentz covariant form as
\bea
\L &=& -\frac{1}{8\pi}\(\frac{1}{12}H^{MNP}H_{MNP} + \frac{N}{\pi^2\sqrt{g}}\frac{g}{6}\epsilon^{\alpha\beta\gamma}H_{\alpha\beta\gamma}  + \frac{N^2}{8\pi^4 g}\sqrt{g}\).
\eea
We then see that we shall take
\bea
\mu eC &=& \frac{N}{2\pi^2\sqrt{g}}.
\eea
With this choice we can match with the expected M five brane Lagrangian. This is a non-trivial consistency check, and we see that we shall interpret this Lagrangian as that it describes a single M five brane coupled to the background gauge potential
\bea
C &=& \frac{N}{2\pi^2 \sqrt{g}} d\sigma^1 \wedge d\sigma^2 \wedge d\sigma^3
\eea
on a three-torus. The $C$-field is large for any finite size of the three-torus when $N$ is large. However in the decompactification limit where the size of the three-torus is taken to infinity, the $C$ field goes to zero.

It is a mystery to us how to extend this to higher values of $k$. It seems that we must take a different background. A three-torus would violate the Dirac charge quantization condition and would instead give $\int H \in 2\pi N \sqrt{k}$ which is unacceptable. 

We note that a similar computation was done \cite{Ho:2008ve}. A difference is that we have no higher order correction terms involving the Nambu bracket in our M five-brane Lagrangian as we take the limit $N\rightarrow \infty$. It is rather finite $N$ that gives a surviving star-product which to lowest order is the Nambu bracket, in the M five theory. This also seems to be in concordance with results from deconstruction of super Yang-Mills theories \cite{Iso:2001mg}. The split $SO(1,5) \rightarrow SO(1,2)\times SO(3)$ was also considered in \cite{Park:2008qe} where the Nambu-Goto action for a five-brane was reformulated in terms of a Nambu bracket square potential defined on three-dimensional space. At the same time auxiliary fields were introduced in order to remove the square root from the Nambu-Goto action.

\section{Non-Abelian extensions}
As was mentioned in the Introduction, the idea of deconstructing a non-Abelian gauge theory is not new. It has appeared in the context of Yang-Mills matter theory in \cite{Iso:2001mg}. An attempt to extend this idea to three-algebra and deconstruct the non-Abelian M five-brane was made in \cite{Ho:2008ve}. Inspired by these works, we consider three-algebra generators on the form
\bea
\T^{aa'}(\sigma,\sigma') &=& \T^a(\sigma)\T^{a'}(\sigma')
\eea
where $\T^a$ and $\T^{a'}$ are functions living on two different three-tori, evaluated at points $\sigma$ and $\sigma'$ respectively\footnote{At this stage we can choose any internal three-manifold for the $\T^{a'}$ and yet obtain an associative three-star-product on $\A\B$. This gives us an abundance of deconstructed non-Abelian theories.}. Now we may consider three different three-algebras. Let us refer to the three-algebra generated by the set $\T^a$ as algebra $\A$, the three-algebra generated by the set $\T^{a'}$ algebra $\B$, and the three-algebra generated by $\T^{aa'}$ algebra $\A\B$. Three-product on $\A$ is different from three-product on $\B$, with a different $\hbar$. Let us denote them as $\hbar_{\A}$ and $\hbar_{\B}$ respectively. The three-product on $\A\B$ is given by
\bea
\T^{aa'} * \T_{cc'} * \T^{bb'} &=& \T^a * \T_c * \T^b \T^{a'} * \T_{c'} * \T^{b'}.
\eea
Associativity of the three-product on $\A\B$ is inherited from associativity of three-products on $\A$ and on $\B$ respectively. We obtain finite-rank gauge structure on the M five-brane if we consider a finite truncation of algebra $\B$. We get a non-commutative M five if we also consider a finite truncation of algebra $\A$. We get an ordinary non-Abelian M five by taking an infinite-dimensional algebra $\A$ and a finite-dimensional algebra $\B$.

The three-bracket of the algebra $\A\B$ is defined as
\bea
[\T^{aa'},\T^{bb'};\T^{cc'}] &=& \T^{aa'}*\T_{cc'}*\T^{bb'} - \T^{bb'}* \T_{cc'} * \T^{aa'}
\eea
and we can compute the contribution up to first order in $\hbar_{\A}$ and all order contributions in $\hbar_{\B}$,
\bea
[\T^{aa'},\T^{bb'};\T^{cc'}] &=& \hbar_{\A} \{\T^a,\T^b,\T_c\}\T^{a'}\T_{c'}\T^{b'} +  \T^a \T_c \T^b [\T^{a'},\T^{b'};\T^{c'}] + \O(\hbar_{\A}^2)
\eea
where, if we would expand, $[\T^{a'},\T^{b'};\T^{c'}] = \hbar_{\B} \{\T^{a'},\T^{b'},\T_{c'}\} + \O(\hbar_{\B}^2)$. However we will not assume that $\hbar_{\B}$ must be very small and will instead work with the exact three-bracket on algebra $\B$. Henceforth $\hbar_{\A}$ and $g_{\A}$ are written shortly as $\hbar$ and $g$. 

Reducing the fuzzy three-torus to a fuzzy two-torus it seems likely that we connect with the number of D2 and number of D4 branes, $N_{\A}$ and $N_{\B}$ respectively. Then we shall take 
\bea
\hbar_{\A} &=& \frac{2\pi}{N_{\A}} \sqrt{g_{\A}},\cr
\hbar_{\B} &=& \frac{2\pi}{N_{\B}} \sqrt{g_{\B}}
\eea
where $g_{\A}$ and $g_{\B}$ denote the determinant of the metric on the respective three-torus. There is no other free parameters in the multiple M five brane theory except from the integer number $N_{\B}$ since we are supposed to take $N_{\A} \rightarrow \infty$. A finite value on $\hbar_{\A}$ corresponds to a non-vanishing non-commutativity parameter in the M five-brane theory. In the M five-brane theory it seems reasonable to think that the dimension of the moduli space or the number of M five-branes coincides with the number of D4 branes to which the theory reduces upon compactifying one longitudinal direction on the M five-branes. The number of M five-branes is then given by $N_{\B}$. The number of three-algebra generators $\T^{a'}$ living on the fuzzy three-torus is $N_{\B}^3$ and this should correspond to the number of degrees of freedom living on the M five-branes.

We make the same rescaling as in the Abelian case to get the M two Lagrangian in the form
\bea
\L &=& \frac{1}{\lambda \hbar} \(\L_m + \L_{CS}\).
\eea
After this rescaling the Nambu-bracket in the Abelian case is replaced by
\bea
\{X^I,X^J;X^K\} &\rightarrow & \{X^I,X^J,X^K\} + \frac{1}{\hbar}[X^I,X^J;X^K]_{\B}.
\eea
The covariant derivative is then
\bea
D_{\mu} X^I &=& \partial_{\mu} X^I + \{X^I,\T^a,\T_b\} A_{\mu}{}^b{}_a + \frac{1}{\hbar}[X^I,\T^{a'};\T^{b'}] A_{\mu}{}^{b'}{}_{a'}
\eea
We define a one-form and a two-form on the M five-brane as
\bea
A_{\mu}(\T^{c'}) &=&  \frac{1}{\hbar}[\T^{c'},\T^{a'};\T^{b'}]A_{\mu}{}^{b'}{}_{a'},\cr
B_{\mu\alpha} &=& -\frac{1}{2\pi\hbar} A_{\mu}{}^b{}_a \T^a \partial_{\alpha} \T_b
\eea
where
\bea
A_{\mu}{}^b{}_a &=& A_{\mu}{}^{bb'}{}_{aa'} \T^{a'} \T_{b'},\cr
A_{\mu}{}^{b'}{}_{a'} &=& A_{\mu}{}^{bb'}{}_{aa'} \T^a \T_b.
\eea
We expand the scalar fields in a similar way as we did in the Abelian case
\bea
X^I &=& T^I + Y^{\alpha}\partial_{\alpha}T^I + Y^A\partial_A T^I.
\eea
The fluctuation fields are now non-Abelian,
\bea
Y^{\alpha} &=& \frac{4\pi^2}{N_{\A}}\frac{g}{2}\epsilon^{\alpha\beta\gamma}B_{\beta\gamma,a'}\T^{a'},\cr
Y^A &=& \frac{4\pi^2}{N_{\A}}\sqrt{g} \phi^A_{a'}\T^{a'}.
\eea
and there are more choices for a background $T^I$. The simplest choice is to take the same $T^I$ as in the Abelian case. In particular then
\bea
\partial_{\alpha'}T^I &=& 0.
\eea
This is a possible choice only if the element '$1$' is an element in the three-algebra $\B$. The general form of the background is 
\bea
T^I &=& T^I_{aa'} \T^a \T^{a'}
\eea
for some constant coefficients $T^I_{aa'}$. It is not immediately clear how we shall choose these coefficients. But if we would not choose $\partial_{\alpha'}\T^I = 0$ we would need to also introduce fields $Y^{\alpha'}$ to be contracted by $\partial_{\alpha'}\T^I$. In this paper we will only consider the case where $T^I$ are exactly the same as in the Abelian case, which means that we require the unit function (a function that takes the constant value one everywhere) belongs to algebra $\B$.

\subsubsection*{Gauge variations}
If we also define gauge parameters out of $\Lambda^{bb'}{}_{aa'}$ similarly to how we defined the gauge potentials,
\bea
-2\pi \Lambda_{\alpha} &=& \Lambda^b{}_a \T^a \partial_{\alpha} \T_b,\\
\Lambda (\T^{c'}) &=& \frac{1}{\hbar}[\T^{c'},\T^{a'};\T^{b'}]\Lambda^{b'}{}_{a'}
\eea
then we get the following induced gauge variations
\bea
\delta B_{\alpha\beta} &=& \partial_{\alpha}\Lambda_{\beta} - \partial_{\beta}\Lambda_{\alpha} - \Lambda (B_{\alpha\beta}),\cr
\delta B_{\mu\alpha} &=& \partial_{\mu}\Lambda_{\alpha} - \partial_{\alpha}\Lambda_{\mu},\cr
\delta A_{\mu} &=& D_{\mu}\Lambda,\cr
\delta \phi^A &=& -\Lambda (\phi^A)
\eea
from (\ref{etta}) and (\ref{tvaa}). Here $D_{\mu}\Lambda = \partial_{\mu}\Lambda + [A_{\mu},\Lambda]$. Also the term $-\partial_{\alpha}\Lambda_{\mu}$ arose as an integration constant. These variations are the exact gauge variations, with no higher order corrections, in the limit $N_{\A} \rightarrow \infty$. 

\subsubsection*{Lagrangian}
We define field strength as
\bea
H_{\mu\alpha\beta} &=& D_{\mu}B_{\alpha\beta} - \partial_{\alpha}B_{\mu\beta} + \partial_{\beta}B_{\mu\alpha}
\eea
and covariant derivatives as
\bea
D_{\mu}\phi^A &=& \partial_{\mu}\phi^A + A_{\mu}(\phi^A),\cr
D_{\mu} B_{\alpha\beta} &=& \partial_{\mu} B_{\alpha\beta} + A_{\mu}(B_{\alpha\beta}).
\eea
When we expand the covariant derivative
\bea
D_{\mu}X^I &=& \partial_{\mu}X^I + [X^I,\T^{aa'};\T^{bb'}] A_{\mu}{}^{bb'}{}_{aa'}
\eea
we get 
\bea
D_{\mu}X^I &=& \frac{4\pi^2}{N_{\A}} \(\frac{g}{2}\epsilon^{\alpha\beta\gamma} H_{\mu\beta\gamma} \partial_{\alpha}T^I + \sqrt{g}D_{\mu}\phi^A \partial_A T^I\).
\eea

We extract the overall coefficient $\frac{1}{(2\pi)^3\lambda\hbar}$ where the $(2\pi)^3$ comes from the measure of the inner product on the three-torus, from the rest of the Lagrangianand. Without this overall coefficient we have the kinetic term
\bea
-\frac{1}{2} \<D_{\mu}X^I,D^{\mu}X^I\> &=& -\frac{1}{2} \(\frac{4\pi^2}{N}\)^2 \(\frac{g}{2} H_{\mu\alpha\beta}H^{\mu\alpha\beta}+gD_{\mu}\phi^A D^{\mu}\phi^A\)
\eea
and the Chern-Simons term
\bea
\L_{CS} &=& (2\pi\hbar)^2 \frac{1}{2} \sqrt{g} \epsilon^{\alpha\beta\gamma}\epsilon^{\mu\nu\lambda}\partial_{\nu}B_{\lambda\alpha}\partial_{\beta}B_{\mu\gamma}\cr
&& +\frac{\hbar}{2} \<[\T^{a'},\T^{b'};\T^{c'}],\T^{d'}\> A^{c'}{}_{b'} d A^{d'}{}_{a'} - \frac{\hbar}{3} \<[\T^{a'},\T^{c'};\T^{d'}],[\T^{b'},\T^{f'};\T^{e'}]\> A^{b'}{}_{a'} A^{d'}{}_{c'} A^{f'}{}_{e'}
\eea
where we here use a rescaled $A^{b'}{}_{a'}$ by a factor of $\hbar$. 

Multiplying back in the overall factor $\frac{1}{(2\pi)^3\lambda\hbar}$, and also including contributions from the sextic potential, we get the full M five-brane Lagrangian 
\bea
\L &=& -\frac{k}{4\pi}\sqrt{g}\(\frac{1}{2} H_{\mu\alpha\beta}H^{\mu\alpha\beta} + \frac{1}{6}H_{\alpha\beta\gamma}H^{\alpha\beta\gamma} - \sqrt{g}\epsilon^{\alpha\beta\gamma}\epsilon^{\mu\nu\lambda}\partial_{\nu}B_{\lambda\alpha}\partial_{\beta}B_{\mu\gamma}\)\cr
&&-\frac{k}{4\pi}\sqrt{g}\(D_{\mu}\phi^A D^{\mu}\phi^A + \partial_{\alpha} \phi^A \partial^{\alpha} \phi^A\)\cr
&&+\frac{N k}{4\pi} \Big(\<[\T^{a'},\T^{b'};\T^{c'}],\T^{d'}\> A^{c'}{}_{b'} d A^{d'}{}_{a'} - \frac{1}{3} \<[\T^{a'},\T^{c'};\T^{d'}],[\T^{b'},\T^{f'};\T^{e'}]\> A^{b'}{}_{a'} A^{d'}{}_{c'} A^{f'}{}_{e'}\Big)\cr
&&+\frac{k}{2}\sqrt{g} \epsilon^{\alpha\beta\gamma}\epsilon^{\delta\epsilon\omega} [B_{\alpha\delta},B_{\beta\gamma};B_{\epsilon\omega}]
\eea
The M-five brane action is given by
\bea
S &=& \int d^3 x \int d^3 \sigma \L
\eea
the measure factor $\sqrt{g}$ sits in $\L$.

The Chern-Simons term in the third line requires some words. First the presence of a factor $N$ could seem to be out of place, especially so since $N$ is taken to be infinite. However if we reduce to a two-torus and use matrices, we find an $1/N$ from the unit normalized trace of matrices, that compensates the factor of $N$ and we see that $k$ is the integer quantized Chern-Simons level. Second we should straighen out the interpretation of the Chern-Simons term above. Quite generally the three-algebra has an associated Lie algebra generated by $[\bullet,T^a;T^b]$. On the Lie algebra we have a Killing form 
\bea
\kappa^a_b,{}^c_d &=& f^{ac}{}_{bd}.
\eea
One can prove that this is a Killing form by deriving the structure constants of the associated Lie algebra from the three-algebra. Then one can show that rising the third Lie algebra index on that structure constants yields a totally antisymmetric structure constant $C^a_b,{}^c_d,{}^e_f = f^{ac}{}_{dg} f^{eg}{}_{bf}$. That means that we have an invariant Killing form. If the Lie algebra is not simple, the Killing form is not unique. However the above Killing form is what enters in the Chern-Simons action. In our case we have
\bea
\<[\T^{aa'},\T^{bb'};\T^{cc'}],\T^{dd'}\> A^{cc'}{}_{bb'} dA^{dd'}{}_{aa'}
\eea
Expanding this out, we find two terms,
\bea
\kappa(A, dA) + \epsilon^{\mu\nu\lambda}\epsilon^{\alpha\beta\gamma} \partial_{\beta}B_{\mu\alpha} \partial_{\nu}B_{\lambda\gamma}.
\eea
Here the first term is given by
\bea
\kappa(A, dA) &=& \int \frac{d^3\sigma}{(2\pi)^3} \{\T^{a'},\T^{b'},\T_{c'}\}\T_{d'} A^{c'}{}_{b'} dA^{d'}{}_{a'}.
\eea
We then note that $\int \frac{d^3\sigma}{(2\pi)^3} \{\T^{a'},\T^{b'},\T_{c'}\}\T_{d'} = f^{a'b'}{}_{c'd'}$ is a Killing form that we denoted by $\kappa$ on the Lie algebra associated with three-algebra $\B$.

We find the presence of a non-dynamical one-form gauge field in the Lagrangian encouraging. It may enable us to construct a reparametrization Wilson surface of the two-form gauge potential by generalizing the construction in \cite{Alvarez:1997ma}. A lattice approach the a non-Abelian Wilson surface can be found in \cite{Rey:2010uz}. It would be very interesting to clarify any possible application of these ideas to our two-form gauge potential, in order to construct the non-Abelian Wilson surface.

\subsection{Cartan sub-three-algebra and degrees of freedom}
For $K=N,N+1,...$ we find three-algebras from the orbifolded fuzzy three-torus which can be realized by $N\times N$ matrices. These correspond to gauge groups $U(N)\times U(N)$. But for $k=1,...,N-1$ we have much bigger three-algebras with the order of $N^3$ three-algebra generators. These have no obvious matrix realizations. 

So far we have assumed that three of the eight transverse coordinates are compact and live on a three-torus. To reduce M theory to type IIA string theory we should to shrink the M theory circle. Let us choose $\sigma^3$ on the three-torus as a coordinate on the M theory circle. Each of the three-torus coordinates run from $0$ to $2\pi$. Three algebra generators are functions on the three-torus. A convenient basis for these generators is
\bea
\T^{\vec{m}} &=& e^{im_{\alpha}\sigma^{\alpha}}
\eea
where 
\bea
m_1,m_2 &=& 0,...,N-1,\cr
m_3 &=& 1, 1 + K,..., 1 + K\[\frac{N-1}{K}\].
\eea
For $K=1$ we have the three-torus. 

We now wish to find a maximal set of three-commuting generators. Let us denote a set of triples $\vec{m}$ that give three-commuting generators as $M$. We will refer to the set 
\bea
\M &=& \{\T^{\vec{m}} | [\T^{\vec{m}},\T^{\vec{n}};\T^{\vec{p}}] = 0
{\mbox{\vspace{1cm} for any $\vec{m},\vec{n},\vec{p}\in M$}}\}
\eea
as the Cartan sub-three-algebra. This shall be a sub-three-algebra with a well-defined dimension. However its elements can not be uniquely chosen. This is a direct generalization of a Cartan subalgebra of a Lie algebra. The condition for three-commuting Cartan generators on a three-torus can be expressed as 
\bea
(\vec{m}\times \vec{n}) \cdot \vec{p} &=& 0.
\eea
This means that all vectors $\vec{m}$ must lie in a two-dimensional plane in $\mb{R}^3$. If $\vec{m},\vec{n},\vec{p}\in M$ then they must lie on the same plane. Say that the plane is described by points $x^{\alpha} \in \mb{R}^3$ satisfying
\bea
(\vec{x} - \vec{a})\cdot \vec{v} &=& 0
\eea
for some displacement vector $\vec{a}$ from the origin, and orthogonal to some vector $\vec{v}\in \mb{R}^3$. From the equality  
\bea
((\vec{m} + \vec{n}-\vec{p}) -\vec{a})\cdot \vec{v} &=& (\vec{m} - \vec{a})\cdot \vec{v} + (\vec{n} - \vec{a})\cdot \vec{v} - (\vec{p} - \vec{a}) \cdot \vec{v}
\eea
it follows that also $\vec{m}+\vec{n}-\vec{p}\in M$. This shows that $\M$ is a sub-three-algebra. 

We may think on $\vec{m}$ as belonging to a cubic lattice in $\mb{R}^3$ with integer spacings. Moreover we have a finite lattice, where each $m_{\alpha} = 0,...,N-1$. We can however embed that lattice in an infinite lattice by imposing periodic boundary conditions. We are interested in planes that intersect a maximal number of lattice points. We believe that any rational plane intersects precisely $N^2$ lattice points, and no more and no less. We can for instance take a plane to span the $m_1$ and $m_2$ directions. This plane will intersect $N^2$ lattice points. 

It is easier to visualize in two dimensions. We can consider some examples that will illustrate the idea quite well. We first take a periodic lattice of points $(m_1,m_2)$ where $m_{\alpha} = 0,1,2,3=N-1$. Thus $N=4$. Let us take a line starting at the point $(0,0)$ and going through the point $(2,3)$. If we count modulo $4$ we now find along this line the following set of points $(0,0)$, $(2,3)$, $(0,2)$, $(2,1)$, $(0,0)$. And from here it repeats itself. The line then goes through four different points. That coincides with the number $N$ in this example. Let us take one more example. Increase to $N=5$ but take again the same line. We then find the points $(0,0)$, $(2,3)$, $(4,1)$, $(1,4)$, $(3,2)$, $(0,0)$,... and we find this time five different points, which again agrees with the number $N$. We expect this to generalize to higher dimensional lattices. In particular for a periodic three-dimensional lattice we expect to find $N^2$ different points along any plane. We have no general proof but we think we have made it plausible enough. 

We conclude that while there are many choices for Cartan-sub-three-algebra generators, its dimension is always the same.

To be specific, let us choose Cartan generators as $T^{(m,n,0)}$ and denote its dimension as dim$(M) = N^2$. 

For any $K\neq 1$ we get $\partial_3 \T^{m} \neq 0$ and the Cartan reduces to $T^{(m,0,1+kn)}$ for $m=0,...,N-1$ and $n=0,...,[(N-1)/K]$. For $k=N,N+1,...$ we must take $n=0$ and we find the dimension of the Cartan is reduced to dim$(M) = N$. For general $K$ we have (we denote by $D$ number of three-algebra generators, and dim$(M)$ dimension of the Cartan)
\bea
{\mbox dim}(M) &=& N \(\[\frac{N-1}{k}\] + 1\),\cr
D &=& N^2 \(\[\frac{N-1}{k}\] + 1\).
\eea
This gives a rather complicated form for $D$ if we express it in terms of dim$(M)$. However things simplify for $k=1$ where we get
\bea
{\mbox dim}(M) &=& N^2,\cr
D &=& N^3
\eea
and for $k=N,N+1,...$ where we get
\bea
{\mbox dim}(M) &=& N,\cr
D &=& N^2.
\eea
This gives an interpolation between dim$(M)^{3/2}$ and dim$(M)^2$ scalings for $K=1$ and $K$ sufficiently large (larger than $N$) respectively. 

This also explains how we can reduce M2 to D2 with gauge group $U(N)$ starting out from three-algebra generators which have $N^3$ components, which in a naive reduction following \cite{Mukhi:2008ux} would give us Lie algebra with $N^3$ generators, as opposed to $N^2$ generators of $U(N)$ gauge group. However the reduction also means that we shall shrink the M theory circle by taking $K$ large. In that reduction we also reduce the number of three-algebra generators from $N^3$ to $N^2$ and no contradiction arises.

\subsubsection*{Further thoughts}
We have proposed a new class of theories for M two-branes when the Chern-Simons level is sufficiently small. More precisely when 
\bea
k\leq N-1.
\eea
It is not entirely clear to us whether these theories are obtainable from ABJM theory. Perhaps the Lagrangian we have proposed is a quantum effective action derived from ABJM theory. If ABJM theory is formulated in terms of three-algebra, then we are still within that framework. What is new is the realization of the three-algebra. 

An important consistency check for any theory that is supposed to describe M two-branes is that it can be reduced to D2 branes. 

One would like to have a substitute for functions and star-three-products, in terms of matrices carrying three indices. This would enable us to reduce M two to D2 for $k=1$ following the same procedure as was done in \cite{Taylor:1996ik} for D-branes. Compactifying one transverse direction of a D-brane leads to T-dual D-brane. Doing the same thing with M two and compactifying one transverse direction, should lead us instead to the D2 brane. It should be noted that this way of reducing M two to D2 is different from what has been done in \cite{Mukhi:2008ux}. Here a vacuum expectation value was given to one of the scalar fields, and at the same time the Chern-Simons level was taken to be large. This is a different kind of reduction altogether. This corresponds to moving the M two branes far away from the orbifold fixed point and letting the orbifold approach the space of a cylinder by taking $k$ large. 

The more interesting way of reducing M two to D2 would be to take $k=1$ and just compactify one transverse direction. 

Attempts have been made to generalize matrices to three-index objects but it does not yet seem to be clear whether this approach can be applied to M two/M five theories. See for instance \cite{Awata:1999dz}, \cite{Rey:2010uz}. One would like to have a three-bracket defined in terms of such matrices carrying three indices.

\subsection{Thoughts on reducing M five to D4}\label{reduction}
By taking $k$ large we can reduce M two to D2 \cite{Mukhi:2008ux}. We can from D2 deconstruct D4. It is then clear that we must also be able to reduce M five to D4 by taking $k$ large. As we have assumed $k=1$ in our deconstruction above, we can not apply this procedure on our deconstructed Lagrangian to reduce it.

We should be able to reduce M two to D2 at $k=1$ by just compactifying one transverse direction on a circle. This has not yet been done in the literature, and we will not attempt this here. 

Dimensional reduction by orbifolding is most clear in the case of a three-sphere embedded in eight-dimensional transverse space. In the large $k$ limit the three-sphere reduces to the two-sphere base manifold. The Abelian D4 on this two-sphere has been deconstructed from ABJM theory in the large $k$ limit in \cite{Nastase:2010uy}.

\section{Three-sphere from ABJM theory}\label{Three-sphere}
The three-torus is a simple geometry, but we find other subtleties instead. We find a background $C$-field, and it is conjectural how to orbifold the background and the three-torus. Thus we computed the deconstructed theory only for $k=1$. 

We can avoid these subtleties by instead considering mass-deformed M two-brane theory. Here we have a maximally supersymmetric three-sphere. Expanding the M two-brane around the three-sphere we can get a single M-five brane wrapped on the three-sphere \cite{Gustavsson:2009qd}. The three-sphere is supersymmetric and we get no bound state of M five and M two-branes. Moreover it is clear how we shall orbifold the three-sphere. If we view the three-sphere as fibered over a two-sphere then we shall orbifold the circle direction.

In the remaining part of this paper we will not go through the deconstruction. The Abelian case has been carried out already to some extent in \cite{Gustavsson:2009qd}. We will be more modest and only show that the fuzzy three-sphere does emerge from mass deformed ABJM theory. At first a three-sphere or rather ${\mb{Z}}_k$ orbifolds thereof was anticipated in mass deformed ABJM theory \cite{Hanaki:2008cu}, but was later shown to be equivalent with a fuzzy (adjoint) two-sphere \cite{Nastase:2010uy}. In this latter reference a D4 brane could be deconstructed on a fuzzy two-sphere in the large $k$ limit. 

Let us consider mass deformed ABJM theory. In a manifestly $SU(4)$ invariant formulation, we have four complex scalar fields $Z^A$. We will split indices as $A = (a,\dot{a})$ and accordingly split $SU(4)\rightarrow SU(2)\times SU(2)$. The scalar fields are expanded as $Z^A = (G^a,G^{\dot{a}})$ plus fluctuations where $G^{\dot{a}}=0$ and $G^a$ satisfy the supersymmetric vacuum equation \cite{Gomis:2008vc}
\bea
G^a &=& [G^a,G^b;G^b].
\eea
We will refer to this equation as the GRVV equation. Irreducible solutions are given by $N\times N$ matrices on the form \cite{Gomis:2008vc}
\bea
G^1 &=& \(\begin{array}{cccc}
0 & 0 & \cdots & 0\\
0 & \sqrt{1} & \cdots & 0\\
  &      &   \ddots & \\
0 &  0   &   \cdots       & \sqrt{N-1}
\end{array}\),\cr
G^2 &=& \(\begin{array}{ccccc}
0 & \sqrt{N-1} & \cdots & 0 & 0\\
0 & 0 & \cdots  & 0 & 0\\
  &   & \ddots &  & \\
0 & 0 & \cdots & 0 & \sqrt{1}\\
0 & 0 & \cdots & 0 & 0.
\end{array}\)
\eea
In particular then 
\bea
G^1 G_1 + G^2 G_2 &=& (N-1)\mb{I}
\eea
which shows that this solution describes a submanifold on a three-sphere. Naively it looks like it could be the full three-sphere. But since $G^1$ is hermitian, $G^1 = G_1$ numerically, this only describes a two-sphere \cite{Nastase:2010uy}.

A slightly modified version of these GRVV matrices was studied in \cite{Kim:2010mr}. One does not have to consider square matrices only since the three-algebra can be realized by $N\times M$ matrices where $M$ and $N$ do not have to be equal. In the next subsection we will see if we can understand what the algebraic GRVV equation describes, if we drop the requirement that it shall be realized by matrices.

\subsection{From GRVV algebra to $SU(2)$ algebra}
It might be of some interest to see that $SU(2)$ algebra can be derived from GRVV algebra. This has been done on the level of matrix realizations in \cite{Nastase:2010uy}, and we will follow this rather closely here, but will work in the three-algebra language and on the level of abstract algebras without assuming any particular realizations thereof. 

Our goal is to eventually answer the question: what does the GRVV equation describe? Showing the emergence of $SU(2)$ should be seen as a first step in that direction. 

Following \cite{Nastase:2010uy}, we let $\sigma_i$ denote the Pauli sigma matrices and we define
\bea
J_i &=& [\bullet,G^{a};G^{b}] (\sigma_i)_{a}{}^{b},\cr
J &=& [\bullet,G^{a};G^{a}].
\eea
We now wish to establish that $J_i$ generate $SU(2)$ and that $J$ commutes with all the $J_i$. 

We can prove the latter statement algebraically just using the GRVV equation in combination with the fundamental identity
\bea
[[X,G^{a};G^{a}],G^{b};G^{c}] - [[X,G^{b};G^{c}],G^{a};G^{a}] &=& [X,[G^{a},G^{b};G^{c}];G^{a}] - [X,G^{a};[G^{a},G^{c};G^{b}]].
\eea
A second form of the fundamental identity reads
\bea
[x,[c,a;b];d] - [x,[c,a;d];b] &=& [x,c;[d,b;a]] - [x,a;[d,b;c]]
\eea
which we can prove by applying the usual form of the fundamental identity twice. We then get
\bea
[X,[G^{a},G^{b};G^{c}];G^{a}] - [X,G^{a};[G^{a},G^{c};G^{b}]] = [X,[G^{a},G^{b};G^{a}];G^{c}] - [X,G^{b};[G^{a},G^{c};G^{a}].
\eea
Finally we use the sphere equation and find this is 
\bea
= -[X,G^{b}; G^{c}] + [X,G^{b}; G^{c}] = 0
\eea
which proves the assertion.

We next wish to prove the $SU(2)$ algebra,
\bea
[J_i,J_j] &=& -2i \epsilon_{ijk} J_k.
\eea
We will approach this in a direct way and yet be completely general by not restricting ourselves to a particular class of realizations of the algebra. We need to obtain the associated Lie algebra of the three-algebra that the sphere equation defines. To this end we should express the sphere equation as a conventional three-algebra (that we may call GRVV algebra),
\bea
[G^1,G^2;G^2] &=& G^1,\cr
[G^2,G^1;G^1] &=& G^2
\eea
and all the other three-brackets vanish, which are not related to the above by antisymmetry such as $[G^2,G^1;G^2] = -[G^1,G^2;G^2]$. Now we also need to express the $J_i$ generators in terms of the three-algebra generators more explicitly,
\bea
J_1(X) &=& [X,G^1;G^2] + [X,G^2;G^1],\cr
J_2(X) &=& i \(-[X,G^1;G^2] + [X,G^2;G^1]\),\cr
J_3(X) &=& [X,G^1;G^1] - [X,G^2;G^2].
\eea
We can then compute 
\bea
J_1(G^1) &=& -G^2,\cr
J_1(G^2) &=& -G^1,\cr
J_2(G^1) &=& -i G^2,\cr
J_2(G^2) &=& i G^1,\cr
J_3(G^1) &=& -G^1,\cr
J_3(G^2) &=& G^2
\eea
and then 
\bea
[J_1,J_2](G^1) &=& 2iG^1,\cr
[J_1,J_2](G^2) &=& -2iG^2,\cr
[J_2,J_3](G^1) &=& 2iG^2,\cr
[J_2,J_3](G^2) &=& 2iG^1,\cr
[J_3,J_1](G^1) &=& -2G^2,\cr
[J_3,J_1](G^2) &=& 2G^1
\eea
Then it is easy to check that 
\bea
[J_i,J_j](G^a) &=& -2i\epsilon_{ijk} J_k(G^a).
\eea
We may note that $J_i$ are hermitan,
\bea
J_i(G_a) &=& \(J_i(G^a)\)^*
\eea
Now we also want to go beyond a $G^a$. But this is easily done by noticing the Leibniz rule,
\bea
J_i(G^a G_b G^c) &=& J_i(G^a) G_b G^c + G^a \(J_i(G^b)\)^* G^c + G^a G_b J_i(G^c).
\eea
Then by noting that if $J_i$ and $J_j$ acts on different elements, they are always arising symmetrically in $i$ and $j$, we see that we only get commutator terms
\bea
[J_i,J_j](G^a G_b G^c) &=& [J_i,J_j](G^a) G_b G^c + G^a \([J_i,J_j](G^b)\)^* G^c + G^a G_b [J_i,J_j](G^c)
\eea
If we can show that $G^a, G^a G_b G^c, ...$ comprise a complete set of three-algebra elements in the sense that they close among themselves under three-bracket then we have established  
\bea
[J_i,J_j] &=& -2i \epsilon_{ijk} J_k.
\eea
It remains to establish the completeness of the set of elements. Let us consider the expansion
\ben
M &=& c_a G^a + c_{a_1 a_2}^{b_1} G^{a_1} G_{b_1} G^{a_2} + \cdots + c_{a_1\cdots a_{\l}}^{b_1 \cdots b_{\l-1}} G^{a_1}G_{b_1}\cdots G_{b_{\l-1}} G^{a_{\l}}\cr
&& + \cdots + c_{a_1\cdots a_{N-1}}^{b_1 \cdots b_{N-2}} G^{a_1}G_{b_1}\cdots G_{b_{N-2}} G^{a_{N-1}}.\label{expansion}
\een
and consider the classical limit where the $G^a$'s are 2-component complex-valued functions. These functions live on some orbifolded sphere $S^3/{\mb{Z}}_k$ for some $k=1,2,3,...,\infty$. When $k=\infty$ we have $S^2 = S^3/{\mb{Z}_{\infty}}$ and we can get a one-to-one map between functions and matrices using star-products. We first ask how many independent components $c_{a_1\cdots a_{\l}}^{b_1 \cdots b_{\l-1}}$ do we have? For commuting functions we find symmetrized indices $a_1\cdots a_{\l}$ as well as $b_1 \cdots b_{\l-1}$. Hence we have $(\l+1)\l$ components. But these are not all independent because of the sphere constraint $G^a G_a = R^2$. For matrices we only need to consider down traces since up traces are related to down traces by the sphere equation. For commuting functions down and up traces are the same, by just commuting the functions. We then need to remove the number of components in a down trace say. There are $\l(\l-1)$ such components. We are left with $2\l$ independent components in $c_{a_1\cdots a_{\l}}^{b_1 \cdots b_{\l-1}}$. Summing them up we get 
\bea
\sum_{\l = 1}^{N-1} 2\l &=& (N-1)N
\eea
components in total. 

Let us now map to the expansion (\ref{expansion}) to bifundamental matrices of gauge group $U(N)\times U(N)$ and keep the same coefficients $c_{a_1\cdots a_{\l}}^{b_1 \cdots b_{\l-1}}$, symmetric and traceless. Since all terms are independent, and they sum up to $(N-1)N$, we have a complete set of generators for $(N-1)\times N$ matrices. Obviously, if $M$, $N$ and $P$ denote any $(N-1)\times N$ matrices, then also $[M,N;P] = MP^{\dag} N - NP^{\dag} M$ must be some $(N-1)\times N$ matrix. That means that the set of $(N-1)\times N$ matrices comprise a complete set of generators of a three-algebra. But as we just have seen, we can express any such $(N-1)\times N$ matrix in the basis given in the expansion (\ref{expansion}). 

We have now completed the proof that the sphere equation {\sl{implies}} that we have $SU(2)$ structure. Thus the GRVV equation can at least describe a fuzzy two-sphere, but hopefully more. 

\subsection{From GRVV algebra to $SO(4)$ algebra}
In \cite{Nastase:2010uy} it was depressingly also shown that the GRVV equation can be {\sl derived} from $SU(2)$, which then would mean that GRVV algebra is isomorphic to $SU(2)$ algebra. That however is a too strong statement as we will see. The proof carried out in \cite{Nastase:2010uy} holds only at the level of matrix realizations of these algebras. 

We can not derive the algebraic GRVV equation from just one $SU(2)$ algebra. To see this it suffices to provide a counter-example. We can not find any counter-example using matrices so let us instead turn to complex valued functions $\G^a$ defined on some yet unspecified three-manifold, and use star-three-product and its associated three-bracket to realize the GRVV equation. Then split these function into real and imaginary parts,
\bea
\G^1 &=& X_1 + i X_2,\cr
\G^2 &=& X_3 + i X_4
\eea
and collectively denote the real coordinates as $X_m$ for $m=1,2,3,4$. Then we see that the GRVV equation, which to leading order in $\hbar$ reads
\bea
\hbar\{\G^a,\G^b,\G_b\} + \O(\hbar^2) &=& \G^a
\eea
can be expressed as
\ben
\hbar\{X_m,X_n,X_p\} + \O(\hbar^2) &=& \epsilon_{mnpq} X_q.\label{threesphere}
\een
This is nothing but the equation of a classical three-sphere. If we define
\bea
J_{mn} &=& \{\bullet,X_m,X_n\} + \O(\hbar)
\eea
then it follows from Eq (\ref{threesphere}) that these generate an $SO(4)$ Lie algbra. Hence they correspond to $6$ Killing vectors on $S^3$. More generally we define $SO(4)$ generators to all orders in $\hbar$ as
\bea
J_{mn} &=& [\bullet,X_m;X_n].
\eea
We have found one $SU(2)$ which is generated by 
\bea
J_i &=& [\bullet,\G^a;\G^b] (\sigma_i)_a{}^b
\eea
We must be able to find one more (commuting) $SU(2)$ in this example since $SO(4) = SU(2) \times SU(2)$. If we represent the sigma matrices so that
\bea
J_1 &=& -[\bullet,\G^1;\G^2]-[\bullet,\G^2;\G^1],\cr
J_2 &=& i[\bullet,\G^1;\G^2]-i[\bullet,\G^2;\G^1],\cr
J_3 &=& [\bullet,\G^1;\G^1]-[\bullet,\G^2;\G^2]
\eea
then we find that the generators in the other $SU(2)$ are given by 
\bea
\t J_1 &=& -[\bullet,\G^1;\G_2]-[\bullet,\G_1;\G^2],\cr
\t J_2 &=& i[\bullet,\G^1;\G_2]+i[\bullet,\G_1;\G^2],\cr
\t J_3 &=& [\bullet,\G^1;\G^1]+[\bullet,\G^2;\G^2].
\eea
In terms of real coordinates we have
\bea
2i J_i &=& \frac{1}{2}\epsilon_{ijk}J_{jk} - J_{i4},\cr
2i \t J_i &=& \frac{1}{2}\epsilon_{ijk}J_{jk} + J_{i4}
\eea
To obtain the $\t J_i$ generators we must have a totally antisymmetric three-bracket. This property does not hold if we realize the three-bracket in terms of matrices and it seems like matrices can not realize the $\t J_i$ generators, but only the $J_i$ generators. This is in agreement with the fact that we can only get one $SU(2)$ or a fuzzy two-sphere if we use matrices to realize the GRVV algebra \cite{Nastase:2010uy}.

Let us view $S^3$ as a circle bundle over $S^2$ with fiber coordinate $\psi \sim \psi + 2\pi$. We can express the embedding coordinates of $S^3 \subset {\mb{C}}^2$ as
\bea
\G^a(\psi) &=& e^{\frac{i\psi}{k}} \t \G^a
\eea
where $\t \G^a$ on the right-hand side are certain functions on the $S^2$ base manifold, obtained by inverting the Hopf map. We will present explicit expressions below. The three-bracket then reduces as
\bea
[\G^a,\G^b;\G^c] &=& [\t \G^a,\t \G^b;\t G^c]_{2}
\eea
where
\bea
[\t \G^a,\t \G^b;\t \G^c]_2 &=& i\(\t \G^a \{\t \G^b,\t \G_c\} - \t \G^b \{\t \G^a,\t \G_c\} - \{\t \G^a,\t \G^b\} \t \G_c\)
\eea
Here
\bea
\{\t \G^a,\t \G^b\} &=& \frac{1}{\sqrt{g_2}} \(\partial_{\sigma} \t \G^1 \partial_{\varphi} \t \G^2 - \partial_{\varphi} \t \G^1 \partial_{\sigma} \t \G^2\)
\eea
is the Poisson bracket on $S^2$ when parametrized by polar coordinates $\sigma,\varphi$, and $g_2 = R^2 \sin \sigma$ is the determinant of the metric $(ds)^2 = R^2\(d\sigma^2 + \sin^2 \sigma d\varphi^2\)$. We note that the three-bracket has lost its total antisymmetry. We can invert the Hopf map as
\ben
\t \G^1 &=& \frac{1}{\sqrt{2}} R \sqrt{1 + \cos \sigma},\cr
\t \G^2 &=& -\frac{1}{\sqrt{2}} R\frac{\sin \sigma}{\sqrt{1+\cos \sigma}} i e^{i \varphi}\label{hopf}
\een
These are functions on $S^2$ base manifold obtained by extracting the phase $e^{i\psi}$ on the fiber. These functions satisfy the following GRVV algebra,
\bea
[\t \G^a,\t \G^b;\t \G^b]_2 &=& \t \G^a.
\eea

For matrices we have the corresponding relation,
\ben
G^a &=& U \t G^a V\label{unitary}
\een
where $U$ and $V$ are unitary matrices ($U^{\dag} = U^{-1}$) and $\t G^a$ now denote the GRVV matrices (which we above denoted as $G^a$. Hopefully this newly introduced tilde in the GRVV matrices does not cause too much confusion). These generate the fiber direction of the fuzzy $S^3$ whereas $\t G^a$ generate the fuzzy $S^2$ base manifold. We notice that for matrices the relation between the two three-brackets reads
\bea
[G^a,G^b;G^c] &=& [\t G^a,\t G^b;\t G^c]_2
\eea
where 
\bea
[\t G^a,\t G^b;\t G^c]_2 &=& U [\t G^a,\t G^b;\t G^c] V.
\eea
One could have wished to find just two complex matrices $G^a$ that generate the whole fuzzy $S^3$, but unfortunately this is impossible \cite{Nastase:2010uy}. We need to consider matrices $U$ and $V$ which can move us along the (fuzzy) fiber direction. 

As thus $[G^a,G^b;G^c] = U [\t G^a,\t G^b;\t G^c] V$ and $G^a = U \t G^a V$, as well as $G^a G_a = \t G^a \t G_a$ if $\t G^a \t G_a \sim \mb{I}$, it follows that both $G^a$ and $\t G^a$ will satisfy the GRVV equation and both will describe spheres. Hence we see that the GRVV equation can describe both a fuzzy three-sphere as well as a fuzzy two-sphere -- the fuzzy two-sphere corresponds to $\t G^a$ and the fuzzy three-sphere corresponds to $G^a$. However so $G^a$ involve $U$ and $V$ matrices and this may not be so nice to use as basic three-algebra generators, and it treats the (fuzzy) fiber direction very differently from the (fuzzy) base-manifold, and we have no concrete matrices to work with, but need plenty of them, one for each choice of $U$ and $V$. While one would have expected to have just four real or two complex matrices to describe the fuzzy three-sphere, just as one has four real embedding coordinates. That however, is impossible. 

\subsection{Convergence of matrices to functions}\label{Convergence of matrices to functions}
So far we have introduced three parameters $\hbar$, $N$ and $R$. It is time to connect these. Let us define
\bea
\G^1 &=& e^{i\varphi} (X^1 + i X^1),\cr
\G^2 &=& e^{i\varphi} (X^3 + i X^4)
\eea
where $X^m$ are real-valued euclidean coordinates of the $S^3$ embedded in $\mb{R}^4$. Hence they are subject to 
\bea
\{X^i,X^j,X^k\} &=& \frac{1}{R^2}\epsilon^{ijkl}X^l,\cr
X^i X^i &=& R^4
\eea
We find it convenient to assume the radius is $R^2$ as the Hopf fibration then gives us radius $R$ of the $S^2$ base manifold. We then get
\bea
\{\G^1,\G^2,\G_2\} &=& -\frac{2e^{2i\varphi}}{R^2} \G^1.
\eea
This leads us to the following identifications,
\bea
\hbar &=& \frac{2}{R^2},\cr
\varphi &=& \frac{\pi}{2}.
\eea

The GRVV matrices give us
\bea
\t G^a \t G_a &=& (N-1)\mb{I}
\eea
while we have for the Hopf projected functions satisfy
\bea
\t \G^a \t \G_a &=& R^2
\eea
This leads us to identify
\bea
R^2 &=& N-1.
\eea
From the GRVV matrices we can obtain
\bea
\tr (\t G^a \t G_b) &=& \frac{(N-1)N}{2}\delta^a_b
\eea
Hence the unit normalized trace form shall be defined as
\bea
\<\t G^a,\t G^b\> &\equiv & \frac{2}{(N-1)N} \tr (\t G^a \t G_b).
\eea
In a similar way we find for the coordinate functions that the unit normalized trace shall be defined as
\bea
\<\t \G^a,\t \G^b\> &\equiv & \frac{1}{2\pi R^3} \int_{S^2} d\Omega_2 \t \G^a \t \G_b
\eea
where $d\Omega_2$ is the volume form of $S^2$ of radius $R$. 

We would now like to establish that a matrix
\bea
M &=& c_a \t G^a + c_{a_1 a_2}^{b_1} \t G^{a_1} \t G_{b_1} \t G^{a_2} + \cdots + c_{a_1\cdots a_{\l}}^{b_1 \cdots b_{\l-1}} \t G^{a_1}\t G_{b_1}\cdots \t G_{b_{\l-1}} \t G^{a_{\l}}\cr
&& + \cdots + c_{a_1\cdots a_{N-1}}^{b_1 \cdots b_{N-2}} \t G^{a_1}\t G_{b_1}\cdots \t G_{b_{N-2}} \t G^{a_{N-1}}
\eea
converges to the function
\bea
\M &=& c_a \t \G^a + c_{a_1 a_2}^{b_1} \t \G^{a_1} \t \G_{b_1} \t \G^{a_2} + \cdots + c_{a_1\cdots a_{\l}}^{b_1 \cdots b_{\l-1}} \t \G^{a_1}\t \G_{b_1} \cdots \t \G_{b_{\l-1}}\t \G^{a_{\l}}\cr
&& + \cdots + c_{a_1\cdots a_{N-1}}^{b_1 \cdots b_{N-2}} \t \G^{a_1}\t \G_{b_1}\cdots \t \G_{b_{N-2}} \t \G^{a_{N-1}}
\eea
as $N\rightarrow \infty$, in the sense that 
\bea
\<M,M\> \rightarrow \<\M,\M\>.
\eea
For any finite $N$ it is crucial we use star-product when multiplying functions. For $N\rightarrow \infty$ the star-product reduces to usual multiplication. 

We will give two examples to illustrate how this can work. First we consider
\bea
\tr (\t G^1 \t G_2 \t G^1 \t G_2 \t G^1 \t G_2) &=& 0
\eea
This is easily seen from the GRVV matrices. We can also see by using (\ref{hopf}) that
\bea
\int d\Omega_2 (\t \G^1)^3 (\t \G_2)^3 &=& 0.
\eea
As our next example we can compute
\bea
\tr (\t G^1 \t G_2 \t G^1 \t G_1 \t G^2 \t G_1 ) &=& \sum_{m=1}^N (m-1)^2(N-m+1)\cr
&=& \frac{N^3}{6}\(1+\O(1/N)\)
\eea
and 
\bea
\int d\Omega_2 (\t \G^1 \t \G_1)^2 \t \G^2 \t \G_2 &=& \frac{8}{3}\pi R^7
\eea
Then we get for the normalized traces,
\bea
\<\t G^1 \t G_2 \t G^1,\t G^1 \t G_2 \t G^1\> &=& \frac{N^2}{6}\(1+\O(1/N)\),\cr
\<\t \G^1 \t \G_2 \t \G^1,\t \G^1 \t \G_2 \t \G^1\> &=& \frac{R^4}{6} 
\eea
By then recalling $R^2 = N (1 + O(1/N))$ we see that these norms indeed agree to leading order in $1/N$.

The order in which matrices are multiplied is irrelevant to leading order as we change the ordering by means of the sphere equation which gives an $1/N$ correction. In the large $N$ limit we have a map from three-commuting matrices, which means diagonal matrices, into functions.
The trace is invariant under unitary transformation 
\bea
\t G^a &\rightarrow & U \t G^a V.
\eea
In the large $N$ limit $\t G^a$ are diagonal and so is $UV$. If we denote its diagonal elements as $e^{i\phi_n}$, this unitary transformation will be mapped into a local phase rotation of functions
\bea
\t \G^a(\sigma,\varphi) & \rightarrow & e^{i\phi(\sigma,\varphi)} \t \G^a(\sigma,\varphi).
\eea

We conclude that we have a fuzzy three-sphere generated by matrices $G^a = U \t G^a V$ for any unitary matrices $U$ and $V$. This is not so nice, since one could have wished for finding a set of only two complex matrices rather than a bunch of them, one for each pair of unitary matrices $U$ and $V$. The best we could do was to describe the fuzzy $S^3$ using two complex matrices $\t G^a$ on the $S^2$ base manifold. This is analogous to describing the classical $S^3$ in terms of functions $e^{i\psi} \t \G^a$ where $\t \G^a$ live on the $S^2$ base manifold and $\psi$ parameterize the fiber. For the functions we can easily imagine functions on $S^3$ being defined as $\G^a = e^{i\psi} \t \G^a$ but there is no corresponding map for matrices from matrices 'valued on' a fuzzy $S^2$ to matrices 'valued on' fuzzy $S^3$. To absorb the degrees of freedom in $U$ and $V$ one could try using three-index objects rather matrices. 

We do not really need to obtain such objects though. It is good enough to just have a star-product of functions. This should be completely equivalent with considering some unknown construction using some kind of generalized matrices, which may not even exist. We can describe the fuzzy $S^3$ using (a finite truncations of) functions on $S^3$ instead of using generalized matrices.  

We note that the three-algebra contains null elements, and any finite-dimensional three-algebra has a matrix realization. So if the number of null elements is finite, then we may hope to find a matrix realization anyway. 

\subsection{Degrees of freedom on fuzzy three-sphere}\label{Degrees of freedom in membrane theories}
As generators in the three-algebra of ABJM theory with gauge group $U(N)\times U(N)$ one may take the set of all bifundamental $N\times N$ matrices $T^a$. As we have seen, any $(N-1)\times N$ matrix can be expanded in the basis (\ref{expansion}) of odd degree monomials built of GRVV matrices $\t G^a$. Let us be ignorant about the discrepance of the missing $N$ components in $(N-1)\times N$ matrices as compared to $N\times N$ as that difference will be subleading in an $1/N$-expansion. 

We may map these matrices to functions $\t \G^a, \t \G^a * \t \G_b * \t \G^c,...$ living on the $S^2$ base manifold and all the three-algebra structure carries over. 

But once we have mapped matrices to functions, we can easily promote these functions to functions $\G^a$ living on the whole $S^3$ or on any orbifold thereof rather than just on the base-manifold. The same GRVV sphere equation still holds after we have promoted $\t \G^a$ to $\G^a$. As the algebra is exactly the same as before, we can use as basis for three-algebra associated to $U(N)\times U(N)$ ABJM theory the functions $\G^a, \G^a * \G_b * \G^c,...$. We then consider the expansion
\bea
\M &=& c_a \G^a + c_{a_1 a_2}^{b_1} \G^{a_1} * \G_{b_1} * \G^{a_2} + \cdots + c_{a_1\cdots a_{\l}}^{b_1 \cdots b_{\l-1}} \G^{a_1} * \G_{b_1} * \cdots * \G_{b_{\l-1}} * \G^{a_{\l}}\cr
&& + \cdots + c_{a_1\cdots a_{N-1}}^{b_1 \cdots b_{N-2}} \G^{a_1} * \G_{b_1} * \cdots * \G_{b_{N-2}} * \G^{a_{N-1}}
\eea
and we want to count the number of independent parameters in this expansion. If we assume that $\G^a$ are coordinate functions on $S^3/{\mb{Z}}_k$ it is plausible to think that these corresponds to the number of degrees of freedom in $U(N)\times U(N)$ ABJM theory at Chern-Simons level $k$. For simplicity let us assume $k=1$. If we counting $c_a$ as $2$ real components, then we count on functions $S^2$ rather than on $S^3$. To count on $S^3$ we must take $c_a$ to have $2$ complex components, or $4$ real components. It is easier to count these components by changing to a real basis with real coordinates $X^m$ for $m=1,2,3,4$. Then the above expansion becomes
\bea
\M &=& d_m X^m + d_{m_1 m_2 m_3} X^{m_1} * X^{m_2} * X^{m_3} + ... + d_{m_1...m_{2N-3}} X^{m_1} * \cdots * X^{m_{2N-3}}.
\eea
The relation between the coefficients $d_{m_1...}$ and the complexified coefficients $c_{a_1..}^{b_1..}$ may be very complicated but in principle it should be possible to work it out by expanding out the functions $\G^a$ in real and imaginary parts, and then multiply them together using star-products. 

The three-sphere constraint in terms of real coordinates reads
\bea
x^m x^m &=& R^4
\eea
and so only traceless coefficients $d_{m_1...}$ are linearly independent. The number of independent components of a symmetric rank $r$ tensor where each index can take $k$ values is given by
\bea
N_{r,k} &=& \frac{(r+1)(r+2)...(r+k-1)}{1...(k-1)}.
\eea
The trace removes two indices so the number of components in the trace is given by $N_{r-2,k}$. In our application we have $k = 4$. Removing trace components of a rank $r$ symmetric tensor leaves us with  
\bea
N_{r,4} - N_{r-2,4} &=& (r+1)^2
\eea
independent components associated to monomial of degree $r$. We shall sum over odd $r$ ranging from $1$ to $2N -3$ to get the total number $D$ of generators in the three-algebra,
\bea
D = \sum_{r = 1}^{2N-2} (2r)^2 = \frac{4}{3}N^3\(1 + \O(1/N)\).
\eea
Let us refer to $D$ as the number of degrees of freedom. 

The dimension of the moduli space corresponds to the number of M2 branes. The moduli space consists of vacuum expectation values of the scalar fields that give vanishing sextic potential. That in turn implies scalar fields for which we have a vanishing three-bracket. Let us consider the north pole 
\bea
x^4 &=& R^2
\eea
on the three-sphere. At the north pole the Nambu bracket is given by 
\bea
\{\T^a,\T^b,\T^c\} &=& \frac{\partial \T^a}{\partial x^1} \frac{\partial \T^b}{\partial x^2} \frac{\partial \T^c}{\partial x^3}.
\eea
The moduli space consist of scalar fields that multiply a restricted set of three-algebra generator, which again consists of odd-degree monomials. Let us restrict to the monomials that do not involve one of the three coordinate $x^{1},x^2$ or $x^3$. If we from our finite list of odd degree monomials remove all monomials that has an explicit dependence on $x^3$, then we find a that these have a vanishing three-bracket. We may say that these can be chosen as Cartan generators of the three-algebra. The Cartan sub-three-algebra is not uniquely determined, and this is just one choice one can make. But the dimension of the Cartan should be independent of the choice we make for the Cartan. The dimension of the Cartan is all that matters to us. Dropping $x^3$ dependence means that the symmetric traceless rank $r$ tensors $d_{m_1...m_r}$ has indices that ranges over only three values instead of four. Apart from that, the counting is exactly the same as before. We first compute
number of independent components in each monomial as
\bea
N_{r,3} - N_{r-2,3} &=& 2r+1
\eea
and then sum them up to the total number of moduli
\bea
M = \sum_{s=1}^{2N-3} 2(2s-1) = 8 N^2 \(1 + \O(1/N)\).
\eea
If we express the number of degrees of freedom $D$ in terms of dimensional of moduli space $M$ we find 
\bea
D &\sim & M^{3/2}.
\eea

\section{Summary and discussion}
We have deconstructed Abelian M five-brane on a three-torus and seen that it matches with the expected Lagrangian coupled to a constant $C$-field. 

We then discussed non-Abelian extensions, but we could find an abundance of extensions. The Lagrangian we presented can not be reduced to super Yang-Mills in any obvious way, one reason for this being that we have fixed the Chern-Simons level to $k=1$ in our computation. We expect that an extension to arbitrary $k$ might be possible to carry out though. This will most likely involve expanding the M two-brane theory about an orbifolded background field configuration. The relation between the Chern-Simons level in ABJM theory, and the three-orbifold needs to be much clarified. 

The choice of background $T^I$ for the scalar fields in the deconstruction is also not clear, and especially so in the non-Abelian case. The choice of auxiliary three-manifold (or orbifold) on which the generators $\T^{a'}$ of algebra $\B$ live, has not been specified. In the large $N_{\B}$ limit such a three-manifold would become smooth and one could think one would deconstruct an M eight-brane. But no such brane exists in M theory. We therefore suspect that one shall not choose the three-manifolds on which $\T^a$ and $\T^{a'}$ live, as different, but as one and the same. That means that we shall consider $\T^a(\sigma)\T^{a'}(\sigma')$ as functions evaluated at two different points on one and the same three-manifold. But this is a speculation. 

The fact that we have failed showing reducibility to D4 puts a possible relation between non-Abelian M five-brane and our non-Abelian extensions on a conjectural status. 

\vskip 2truecm

{\sl{Acknowledgements}}:
I would like to thank Soo-Jong Rey for many helpful discussions, correspondences and encouragement. Early on when this work was at a preliminary stage, Jeong-Hyuck Park and Kanghoon Lee helped me check explicitly that the naive definition of the star-three-product is not associative and does not give a three-bracket that satisfies the fundamental identity. This work was supported by the National Research Foundation of Korea(NRF) grant funded by the Korea government(MEST) through the Center for Quantum Spacetime(CQUeST) of Sogang University with grant number 2005-0049409.

\newpage
\appendix
\section{Totally antisymmetric three-bracket}
Here we will show that the three-bracket 
\bea
[\T^a,\T^b;\T^c] &=& \T^a * \T_c * \T^b - \T^b * \T_c * \T^a
\eea
is totally antisymmetric. 

The antisymmetry is manifest at linear order, where we have 
\bea
\hbar \{\T^a,\T^b,\T_c\} = -\hbar\{\T^a,\T_c,\T^b\}.
\eea

To show the antisymmetry property in general, let us choose Riemann normal coordinates locally around a point $\sigma^{\alpha}$ in the three-manifold $M$ at which we are interested to compute the star-product. Then we have in that point 
\bea
g_{\alpha\beta} &=& \delta_{\alpha\beta},\cr
\partial_{\gamma}g_{\alpha\beta} &=& 0.
\eea
We may on a local patch around that point choose a Fourier basis
\bea
\T^a(\sigma) &=& \sum_k e^{ik_{\alpha}\sigma^{\alpha}} \T^a_k
\eea
The allowed values of $k$ depends on the global structure of $M$, or on how various patches are glued together. We define
\bea
\T^a_k &=& \T^{a,-k}.
\eea
We now have
\bea
\T^a * \T_c * \T^b &=& \exp \frac{i\hbar}{2} \{k,k',k''\} \T^a_k \T_c^{k''} \T^b_{k'} e^{i(k+k'-k'')\sigma}.
\eea
Then 
\bea
[\T^a,\T^b;\T^c] &=& \(\exp \frac{i\hbar}{2} \{k,k',k''\} - \exp -\frac{i\hbar}{2} \{k,k',k''\}\) \T^a_k \T_c^{k''} \T^b_{k'} e^{i(k+k'-k'')\sigma}.
\eea
We now also have
\bea
[\T^a,\T_c;\T_b] &=& \(\exp \frac{i\hbar}{2} \{k,k',k''\} - \exp -\frac{i\hbar}{2} \{k,k',k''\}\) \T^a_k \T^{b,k''} \T_{c,k'} e^{i(k+k'-k'')\sigma}\cr
&=& \(\exp \frac{i\hbar}{2} \{k,k',k''\} - \exp -\frac{i\hbar}{2} \{k,k',k''\}\) \T^a_k \T^{b}_{-k''} \T_{c}^{-k'} e^{i(k+k'-k'')\sigma}\cr
&=& \(\exp \frac{i\hbar}{2} \{k,k',k''\} - \exp -\frac{i\hbar}{2} \{k,k',k''\}\) \T^a_k \T^{b}_{k''} \T_{c}^{k'} e^{i(k-k'+k'')\sigma}\cr
&=& \(\exp \frac{i\hbar}{2} \{k,k',k''\} - \exp -\frac{i\hbar}{2} \{k,k',k''\}\) \T^a_k \T_{c}^{k'} \T^{b}_{k''} e^{i(k+k''-k')\sigma}\cr
&=& -[\T^a,\T^b;\T^c].
\eea

\section{Associativity of the star-three-product}
We have not found any short proof of associativity. We simply start by computing
\bea
((\T^a * \T_c * \T^b)* \T_e * \T^d)(\sigma) &=& \lim_{\sigma^{out}\rightarrow \sigma} \lim_{\sigma''''\rightarrow \sigma'''=\sigma^{out}} \cr
&&\exp\frac{\hbar}{2}\{\bullet,\bullet'''',\bullet'''\}\cr
&&\lim_{\sigma'\rightarrow \sigma} \lim_{\sigma''\rightarrow \sigma'}\cr
&&\exp\frac{\hbar}{2}\(\{\bullet,\bullet'',\bullet'\} + \{\bullet'',\bullet',\bullet^{out}\} - \{\bullet,\bullet',\bullet^{out}\} - \{\bullet,\bullet'',\bullet^{out}\}\)\cr
&&\T^a_{k} \T_{c,k'} \T^b_{k''} \T_{e,k'''} \T^d_{k''''} e^{ik\sigma} e^{ik'\sigma'} e^{ik''\sigma''} e^{ik'''\sigma'''} e^{ik''''\sigma''''}
\eea
We find the result
\bea
&&\exp\frac{i\hbar}{2}\{k+k'+k'',k'''',k'''\} \cr
&& \exp\frac{i\hbar}{2}\(\{k,k'',k'\} + \{k'',k',k'''+k''''\} - \{k,k',k'''+k''''\} - \{k,k'',k'''+k''''\}\)\cr
&&\T^a_{k} \T_{c,k'} \T^b_{k''} \T_{e,k'''} \T^d_{k''''} e^{i(k+k'+k''+k'''+k'''')\sigma}
\eea
and we see the emergence of $10$ terms if we expand out the Nambu brackets. What these Nambu brackets mean is just
\bea
\{k,k',k''\} &=& \epsilon^{\alpha\beta\gamma}k_{\alpha}k'_{\beta}k''_{\gamma}.
\eea
If we compute $(\T^a * (\T_c * \T^b* \T_e) * \T^d)(\sigma)$ we get the result
\bea
&&\exp\frac{i\hbar}{2}\{k,k'''',k'+k''+k'''\} \cr
&& \exp\frac{i\hbar}{2}\(\{k',k''',k''\} + \{k''',k'',k''''+k\} - \{k',k'',k''''+k\} - \{k',k''',k''''+k\}\)\cr
&&\T^a_{k} \T_{c,k'} \T^b_{k''} \T_{e,k'''} \T^d_{k''''} e^{i(k+k'+k''+k'''+k'''')\sigma}
\eea
and if we compute $(\T^a * \T_c * (\T^b* \T_e * \T^d))(\sigma)$ we get the result
\bea
&&\exp \frac{i\hbar}{2}\{k,k''+k'''+k'''',k'\}\cr
&&\exp \frac{i\hbar}{2}\(\{k'',k'''',k'''\}+\{k'''',k''',k+k'\}-\{k'',k''',k+k'\}-\{k'',k'''',k+k'\}\)\cr
&&\T^a_{k} \T_{c,k'} \T^b_{k''} \T_{e,k'''} \T^d_{k''''} e^{i(k+k'+k''+k'''+k'''')\sigma}
\eea
Expanding out the antisymmetric brackets, and placing the number of primes in increasing order we find in both cases a sequence of ten terms all coming with a minus sign as
\bea
-\{k,k',k''\}-\{k',k'',k'''\}-...-\{k'',k''',k''''\}
\eea
and so the two diffferent ways of multiplying together the elements give the same answer. 

It is slightly non-trivial to see that this associativity extends when we consider multiple star-products since the $\bullet^{out}$ acts on all the outer star-products. Let us check that
\bea
((\T^a * \T_c * \T^b) * \T_e * \T^d) * \T_g * \T^f &=& (\T^a * \T_c * \T^b) * (\T_e * \T^d * \T_d) * \T^f
\eea
For the left-hand side we get
\bea
&&\exp \frac{i\hbar}{2} \Big(\{k^0,k^2,k^1\} + \{k^2,k^1,k^3+k^4+k^5+k^6\} \cr
&&-\{k^0,k^1,k^3+k^4+k^5+k^6\} - \{k^0,k^2,k^3+k^4+k^5+k^6\}\Big)\cr
&&\times \exp \frac{i\hbar}{2} \Big(\{k^0+k^1+k^2,k^4,k^3\} + \{k^4,k^3,k^5+k^6\} \cr
&&- \{k^0+k^1+k^2,k^3,k^5+k^6\} - \{k^0+k^1+k^2,k^4,k^5+k^6\}\Big)\cr
&&\times \exp \frac{i\hbar}{2} \{k^0+k^1+k^3+k^4,k^6,k^5\}
\eea
and for the right-hand side we get
\bea
&&\exp \frac{i\hbar}{2} \Big(\{k^0,k^2,k^1\} + \{k^2,k^1,k^3+k^4+k^5+k^6\} \cr
&&- \{k^0,k^1,k^3+k^4+k^5+k^6\} - \{k^0,k^2,k^3+k^4+k^5+k^6\}\Big)\cr
&&\times\exp \frac{i\hbar}{2} \Big(\{k^3,k^5,k^4\} + \{k^5,k^4,k^0+k^1+k^6\} \cr
&&-\{k^3,k^4,k^0+k^1+k^2+k^6\} - \{k^3,k^5,k^0+k^1+k^2+k^6\}\Big)\cr
&&\times\exp \frac{i\hbar}{2} \{k^0+k^1+k^2,k^5,k^3+k^4+k^5\}.
\eea
Expanding these expressions, we find in both left-hand side and right-hand side $35$ terms, which agrees with the total number of permutations one can have, and if we arrange the $k^i$'s in increasing order we find that each terms comes with a minus sign on both sides. 

We can argue that associativity for multiple three-products follows by induction. If for instance in the above example we substitute $\T^a* \T_c * \T^b$ with $T^h$, then the relation we just showed above reads
\ben
(\T^h * \T_e * \T^d) * \T_g * \T^f &=& \T^h * (\T_e * \T^d * \T_g) * \T^f\label{assocrel}
\een
and so it seems we can proceed iteratively to prove associativity for expressions involving an arbitrary number of three-products. 

However things are not quite that simple since we have shown associativity only when $T^h$ is a function, but here $\T^h = \T^a* \T_c * \T^b$ contains also a differential operator acting on outer three-products. More precisely a differential operator which acts on $\T_e T^d \T_g \T^f$ where we have usual multiplication of four functions. Multiplication of functions at the same point is associative. Acting on that product of four functions by a differential operator still gives an associative product. Hence the fact that $T^h$ is now extended to a differential operator does not violate the associativity relation (\ref{assocrel}), and we can therefore prove associativity by induction.

\end{document}